\documentclass[useAMS,usenatbib,letter]{mn2e}
\usepackage{times}
\usepackage{graphicx}
\usepackage{amssymb}
\usepackage{natbib}
\usepackage{color,ulem}
\bibpunct{(}{)}{;}{a}{}{,}

\title[Galaxy populations in Antlia. I.]{Galaxy populations
in the Antlia cluster. I.\,Photometric properties of early-type galaxies 
\thanks{This paper is based on data obtained with the 4m telescope at CTIO,
  Chile, with the 6.5 meter Magellan telescopes at Las Campanas
  Observatory, Chile, and at CASLEO, operated under agreement 
  between CONICET and the Universities of La Plata, C\'ordoba and 
  San Juan, Argentina.}  }
\author[Smith Castelli et al.]{Anal\'{\i}a V. Smith Castelli$^{1}$ 
\thanks{E-mail:\,asmith@fcaglp.unlp.edu.ar\,\,(ASC);\,lbassino@fcaglp.unlp.edu.ar
\,(LB);\,tom@mobydick.cfm.udec.cl\,\,(TR);\,scellone@fcaglp.unlp.edu.ar\,\,(SC);
\,linfante@astro.puc.cl\,\,(LI)},
Lilia P. Bassino$^{1}$, Tom Richtler$^{2}$, Sergio A. Cellone$^{1}$, 
\newauthor Cristian Aruta\thanks{\it In memoriam} and Leopoldo Infante$^{3}$\\ 
$^{1}$Facultad de Ciencias Astron\'omicas y Geof\'{\i}sicas,
      Universidad Nacional de La Plata,
      Paseo del Bosque, B1900FWA La Plata,
      Argentina;\\ and IALP (CONICET-UNLP) \\
$^{2}$Departamento de F\'isica, Universidad de Concepci\'on, Casilla 160-C, 
Concepci\'on, Chile \\
$^{3}$Departamento de Astronom\'ia y Astrof\'isica, Pontificia Universidad 
Cat\'olica de Chile, Casilla 306, Santiago 22, Chile} 

\begin{document}

\date{Accepted . Received ; in original form }
\maketitle

\label{firstpage}

\begin{abstract}
We present the first colour-magnitude relation (CMR) of early-type galaxies
in the central region of the Antlia cluster, obtained from CCD wide-field
photometry in the Washington photometric system.  Integrated $(C-T_1)$
colours, $T_1$ magnitudes, and effective radii have been measured for 93
galaxies (i.e. the largest galaxies sample in the Washington system till now)
from the FS90 catalogue \citep{FS90}. Membership of 37 objects can
be confirmed through new radial velocities and data collected from the
literature. The resulting colour-magnitude diagram shows that early-type
FS90 galaxies that are spectroscopically confirmed Antlia members or that
were considered as definite members by FS90, follow a well defined CMR 
($\sigma_{(C-T_1)}\sim$ 0.07\,mag) that spans 9 magnitudes 
in brightness with no apparent change of slope. This relation is very 
tight for the whole magnitude range but S0 galaxies show a larger dispersion, 
apparently due to a separation of ellipticals and S0s. Antlia displays a slope 
of $-$ 13.6 in a $T_1$ vs. $(C-T_1)$ diagram, in agreement with results for 
clusters like Fornax, Virgo, Coma and Perseus, which are dynamically 
different to Antlia. This fact might indicate that the build up of the CMR in
cluster of galaxies is more related to galaxies internal processes than to
the influence of the environment. Interpreting the CMR as a
luminosity-metallicity relation of old stellar systems, 
the metallicities of the Antlia galaxies
define a global relation down to $M_{\rm V}\approx-$ 13. 
We also find, for early-type dwarfs, no clear relation between luminosity and
effective radius, indicating a nearly constant mean effective radius of 
$\sim 1$ kpc. This value is also found in several samples of dwarf galaxies
in Virgo and Coma. 
\end{abstract}

\begin{keywords}
galaxies: clusters: general -- galaxies: clusters: individual: Antlia -- 
galaxies: elliptical and lenticular, CD -- galaxies: dwarf -- 
galaxies: photometry 
\end{keywords}
\section{Introduction}
\label{intro}

It is well known that early-type galaxies in clusters and groups define a
tight sequence in the colour-magnitude diagram (CMD), in the sense that more
luminous ellipticals (Es) are redder than fainter ones
(\citealp*[e.g.][]{B59,dV61,VS77,B92,C06}). Many spectroscopic studies of
giant ellipticals (e.g. \citealt[][for the Fornax cluster]{Ku00};
\citealt[][for the Virgo cluster]{Va01}; see also \citealt{TF02,Ch06}) and
dwarf galaxies (e.g. \citealt{C02,T03a}; \citealt*{T03b,VZ04};
\citealt{M07}) have shown that the colour-magnitude relation (CMR) mainly
reflects metallicity effects. A reasonable explanation, that comes from
an analysis of the CMR, is based on the assumption that
the more luminous (massive) galaxies, capable of retaining
their metal content due to their deep potential wells, can be enriched to
higher levels than low-mass Es which are more sensitive to the effect of
mass loss by galactic winds and supernovae (\citealt{D84,K97};
\citealt[][and references therein]{R01}). In a cluster environment, tidal
stripping of gas and/or interaction with the intracluster medium might also
play a role. See \citet{derijke05} for an account of theoretical models.

Although also early-type dwarf galaxies in clusters and groups seem to
follow a well defined CMR (\citealp{C83}; \citealp*{S97,H03,L04}; 
 \citealp{C06,M07}; \citealp*{LGB08}), it is not yet clear whether 
this relation broadens
towards fainter magnitudes \citep*{Con02,Con03} or just extends the one
followed by the Es \citep{Ada06} with a similar slope and level of
scatter. \citet{B92} report that the CMR defined by the luminous early-type
galaxies has the same form in the Virgo and Coma clusters, even using
different colours ($U-V$, $V-K$ or $J-K$) vs. the total $V$ magnitude. 
\citet{LGB05} \citep[see also][]{LGB08} have shown that nucleated dEs in Virgo 
follow a tight CMR with no broadening towards fainter magnitudes, which extends 
to brighter Es but with a change in slope. Furthermore, 
these authors point out that dS0 should be considered as a separate class of dwarf 
galaxies in order to properly analyse CMDs.

\citet*{T01} perform an $(U,V)$ photometric study of galaxies in the Coma 
cluster to investigate the dependence of the slope and dispersion of the CMR  
on the morphology and luminosity of the galaxies, and on their 
environmental properties within the cluster. They find the CMR 
to be consistent when compared between samples of early-type galaxies with 
different characteristics. In particular, no variation in the CMR slope 
between E and S0 galaxies is detected. 

However, a common and well defined CMR of early-type galaxies over a broad 
magnitude range seems to be surprising, given that the chemical histories 
of giant ellipticals and dwarf galaxies have supposedly been quite different. 
The $\alpha$-element overabundance of giant ellipticals points to post-star 
burst populations, rapidly enriched by type II-supernovae, while the subsolar 
$\alpha$-abundances of dwarf galaxies rather speak for a continuous
enrichment, which has ceased because of gas removal by galactic outflows or 
stripping (\citealp{T03b,VZ04}; for an alternative view, see also 
\citealp*{KWK07}). 

In this context, nearby galaxy clusters are obviously of particular
interest.  We have started the Antlia Cluster Project whose goal is to study
the galaxy population of the Antlia cluster, which is the third nearest well
populated galaxy cluster after those of Virgo and Fornax.  Despite its
proximity, richness and concentration, Antlia is until now practically
unexplored. It is located between the third and fourth Galactic quadrants,
not too far from the Galactic plane (l~$\approx~270\degr$, b~$\approx~20\degr$). 

Antlia exhibits a complex structure consisting of several subgroups, the
most conspicuous ones being dominated by the giant elliptical galaxies 
NGC\,3258 and NGC\,3268. X-ray observations showed extended emission around 
both subgroups \citep*{ped97,nak00}. The gravitating masses estimated in these 
studies for each group are of the order of 1.9 $\times 10^{13} 
\rm{M}_{\odot}$ within a radius of $\sim 250$ kpc, i.e. similar to what is 
found in Fornax within the same radius. In both subgroups, the emission is 
concentrated towards the dominant galaxy, but extensions elongated in the 
direction to the other subgroup are also present. This kind of substructure 
depicted in X-rays suggests an ongoing merger.

Here, we shall adopt the distance modulus $(m-M)=32.73$ given by \citet*{dir03}, 
which was calculated as the mean of the distances towards the two giant 
ellipticals obtained by \citet{pru96} and \citet*{ton01}.
According to the distance moduli estimated for 
both giants with the surface brightness fluctuation (SBF) method by 
Tonry et al., they are separated by about 3 Mpc, but this difference 
is not conclusive because their distance moduli as well as their radial 
velocities agree within the errors.

\citet{dir03}, performed the first 
investigation of the globular cluster (GC) systems around NGC\,3258 and NGC\,3268. 
They showed that both cluster systems are 
elongated in the same direction as a connecting line between the two galaxies, 
resembling the X-ray results. More recent results on these GC systems can be 
found in \citet{har06} and \citet*{bas07}, who show that
the GC systems colour distributions are bimodal but the brightest GCs
present a unimodal distribution. Furthermore, the red (metal-rich) GCs
follow closely the galaxies' surface brightness profiles, and the
estimated total GC populations are $6000 \pm 150$ GCs for NGC\,3258 and
$4750 \pm 150$ GCs for NGC\,3268.

The photographic work of \cite{FS90} was the first and last major effort
devoted to study the galaxy population of the Antlia cluster. They
identified, by visual inspection of photographic plates, 375 galaxies that
are listed in their Antlia Group Catalogue (named hereafter with the acronym
FS90 plus the catalogue number). It gives, among other data, a membership
status (1: definite member, 2: likely member, 3: possible member) and a
morphological type for each galaxy. The membership status was mainly 
based on morphological criteria, i.e. surface brightness, resolution into 
knots and late-type galaxies' luminosity class 
\citep*[see][for more details]{B85}, as only about 6\% of the galaxies in 
this catalogue had available radial velocities. 
They showed that the central galaxy
density in Antlia is a factor 1.4 higher than in Fornax, and 1.7 higher than
in Virgo.
  
In this first paper of the Antlia Cluster Project, we present initial
results from a photometric study, using the Washington photometric system 
\citep{C76}, of 100 FS90 galaxies in the central region of
the cluster. To our knowledge, this galaxy sample is the largest one studied  
with the Washington system till now. We also add some new radial velocities 
for several galaxies. In Section\,\ref{observations}, we give information about 
the photometric and spectroscopic observations, as well as the data. In
Section\,\ref{photometry}, we show the CMD and the surface brightness -
luminosity diagram for 93 FS90 galaxies placed in the central region of 
Antlia.  We present a discussion of our work in Section\,\ref{discussion} and, 
in Section\,\ref{conclusions}, our conclusions.

\section{Data and Reduction}
\label{observations}

\begin{figure}
\includegraphics[width=84mm]{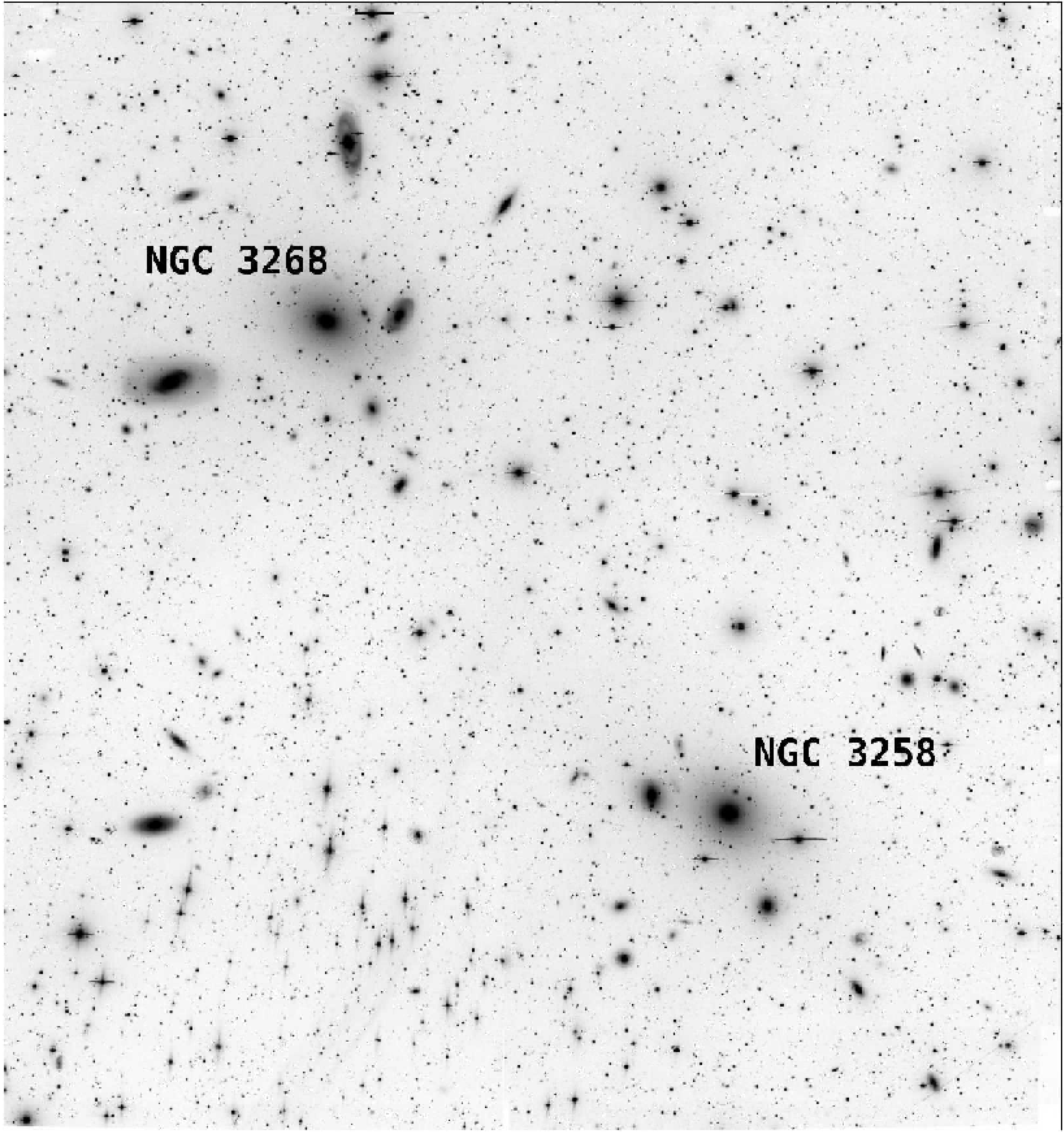}
\caption{$C + R $ combined image of the MOSAIC field. Labels indicate
the two dominant galaxies of the Antlia cluster. North is up and East 
to the left.}
\label{mosaic}
\end{figure}

The photometric observations were performed with the MOSAIC camera (8 CCDs
mosaic imager) mounted at the prime focus of the 4-m Blanco telescope at the
Cerro Tololo Inter--American Observatory (CTIO), during 2002 April 4--5.
One pixel of the MOSAIC wide-field camera subtends 0.27 arcsec on the sky,
which results in a field of $36 \times 36$ arcmin$^2$, about 
$370 \times 370$ kpc$^2$ at the Antlia distance (35.2 Mpc). The central part 
of the cluster has been
covered by one MOSAIC field (Fig\,\ref{mosaic}), with both short and long 
exposures. The same material has already been used to investigate the GC 
systems of the dominant elliptical galaxies. Details are given in 
\citet{dir03}.

Kron-Cousins $R$ and Washington $C$ filters were used. We selected the $R$
filter instead of the original Washington $T_1$ as \citet{gei96} has shown
that the Kron-Cousins $R$ filter is more efficient than $T_1$, as it has 
a better transmission at all wavelengths, and that $R$  
and $T_1$ magnitudes are very similar, with just a very small colour term
and zero-point difference ($R - T_1 \approx -0.02$). The seeing on
the $R$ image is 1$''$ and on the $C$ image is 1.1$''$.

Medium-resolution spectroscopic observations were performed during two
nights (2004 January 19-20) with the IMACS areal camera and spectrograph
mounted on the Magellan I Baade 6.5-m Telescope (Las Campanas Observatory,
Chile).  Additional low-resolution spectra for three FS90 blue compact dwarf
(BCD) candidates were obtained during the nights of March 12-13, 2007, with
the REOSC spectrograph at the ``Jorge Sahade'' 2.15\,m telescope of La Plata
University, Argentina. Radial velocities were measured by cross-correlation
and, in the case of BCDs, by fitting bright emission lines. 
Details will be given in Smith Castelli et al. (in preparation).

A total of 100 galaxies from the FS90 Antlia Group Catalogue are located in
our field. The brightest ones are overexposed in our long-exposure 
frames. To obtain colours and magnitudes for the non-overexposed FS90
galaxies, we ran SExtractor \citep{ba96} on our long-exposure $R$ frame. 
We considered as detection criteria 10 connected pixels above 1.5$\sigma$ of 
the background level. A pyramidal filter was applied and the global 
background map used for detection was constructed by setting a mesh size of 64 
$\times$ 64 pixels$^2$ ($\sim$ 17 $\times$ 17 arcsec$^2$). To obtain accurate
photometry, we used the local background option of SExtractor by 
setting rectangular annuli with widths of 24 pixels around the objects 
\citep{ba96}. We adopted MAG\_AUTO, i.e. a Kron-like elliptical
aperture magnitude, as a representative total magnitude of 
the detected objects in both filters \citep{non99}. To get $(C-T_1)$ colours 
in a consistent manner, the corresponding $C$ magnitudes were measured using 
the same coordinates and Kron-like elliptical apertures obtained from the $R$ 
image. 

In order to obtain integrated magnitudes and colours for faint objects that 
were not properly detected by SExtractor, as well as for the brightest 
galaxies whose centres are overexposed on the long-exposure frames, we worked 
with the task ELLIPSE within IRAF (more details will be given in Smith Castelli
et al., in preparation). In all these cases, we used smaller trimmed images in 
both filters, containing the target galaxies, and measured the background 
locally. Note that in the case of the brightest galaxies, the use of MAG\_AUTO 
would have lead to a severe underestimation of the light of the galaxy 
\citep{GD05}. For fainter objects (i.e. those
displaying rather exponential profiles), however, total magnitudes obtained 
from Kron-like elliptical apertures can be considered as representative of the 
total luminosity of the galaxy, as stated by Graham \& Driver. 

As said above, for each faint object not detected by SExtractor
we have used trimmed and registered images 
of the same sizes in $R$ and $C$ filters. We estimated the sky level  
from the statistic provided by IMEXA. After subtracting the sky level and 
masking foreground stars, we run ELLIPSE on the $R$ image. The photometry of 
the sky subtracted $C$ image was obtained considering the same elliptical 
apertures of the $R$ image with the aim at estimating consistent colours. 
Once we obtained the photometry in both filters, we corrected the sky level
by constructing growth curves, i.e. we plotted the integrated flux within the 
elliptical apertures vs. the semi-major axis of the apertures. The correction 
to the sky level of an image is the value for which these curves display an 
asymptotic flat behaviour to infinity. Therefore, we added or subtracted 
different constant values (smaller than 2\% of the original estimated level) 
to the integrated fluxes until we got asymptotically flat curves. These 
corrected fluxes provided the brightness profiles that were numerically 
integrated to obtain the integrated magnitudes in each filter. From these 
magnitudes we calculated the integrated colours.

For the galaxies with overexposed centres, 
we considered long and short-exposure images in both 
filters. We rescaled the intensities of the 
short-exposure images to the long-exposure ones, performed internal 
fits on the sky subtracted short-exposure $R$ images and external fits on   
the sky subtracted long-exposure ones. We then used the corresponding  
elliptical apertures to perform the photometry on the sky subtracted short 
and long-exposure $C$ images. To obtain the whole brightness profile in each 
filter, we merged  the internal and external fits, and we corrected the 
external sky level as it had been calculated for faint objects. Again, 
the corrected 
fluxes provided the brightness profiles that were numerically integrated to 
obtain total magnitudes and colours.

Magnitude and colour errors were estimated using the equations given 
by \citet{LGB08} in their section 4, but omitting the last term in equation
(3) (we note that this equation refers to a relative flux error). In our 
case, the uncertainty in the sky level determination per pixel is taken
to be 8\%, as it is obtained from our images. Following Lisker et al.,
we also consider the flux uncertainty caused by the uncertainty 
in the semi-major axis of our elliptical 
apertures to be similar to that of the sky level. 

The FS90 number, NGC number, J2000 coordinates, FS90 morphology and
membership status, and $E(B-V)$ values for 100 FS90 galaxies are listed in
Table\,\ref{tabla} in the first 7 columns. The eighth and ninth columns give
the total magnitude and colour (not corrected by absorption or reddening),
respectively, with errors in parenthesis. The tenth column gives 
the surface brightness of the limiting
isophote, within which the total magnitude has been calculated. The eleventh
column lists the corresponding limiting radius. The twelfth column gives the
mean surface brightness within the effective radius (i.e. the radius
containing half of the light), and the thirteenth column the effective
radius measured in the $R$ band. The fourteenth column gives the radial 
velocities derived from our spectra, as well as those obtained from 
NED\footnote{This research has made use of the NASA/IPAC Extragalactic 
Database (NED) which is operated by the Jet Propulsion Laboratory, California 
Institute of Technology, under contract with the National Aeronautics and 
Space Administration}, and the last column shows some remarks.

Given that our sample covers a large luminosity range, it is interesting 
to estimate the signal-to-noise ratio (S/N) of one of the faintest and one of
the brightest galaxies. For the $R$ filter we have obtained an S/N within 
$r_{\rm T_1}$ (as defined in the Notes to Table\,\ref{tabla}) of about 100 for 
FS90\,114, and S/N $\sim$ 16000 for FS90\,185 (NGC\,3269). The S/N values of 
the $C$ images are between one half and one third of the $R$ images.

Taking as a reference the mean radial velocities of Antlia early- and 
late-type galaxies given by \citet{hop85} and considering a 
dispersion of 3$\sigma$ from these values, we will consider as Antlia members 
all objects with velocities in the range 1200 - 4200 km\,s$^{-1}$. We prefer 
to take  relaxed membership criteria given the substructure of the Antlia 
cluster and its probably complex dynamics. However, we note that, 
with the exception of four S0 galaxies with radial velocities below 
2000 km\,s$^{-1}$, early-type galaxies with known radial velocities are 
confined in the range 2400 - 3900 km\,s$^{-1}$. We point out that 14 objects 
had radial velocities both from the literature and from our 
spectroscopic data. Our values are in quite good agreement with previous 
measurements as we obtain $\langle |V_{\rm lit} - V_{\rm ours}| \rangle
=63.5\pm55.9$ km s$^{-1}$.

Six galaxies out of the 100 FS90 
objects present in our field have no photometric information: FS90\,169 
is located within a gap of the image, FS90\,223 is affected by strong bleeding,
FS90\,148 is extremely faint and placed 
near a star, and three galaxies (namely, FS90\,121, FS90\,178 and 
FS90\,235) have doubtful coordinates. Also FS90\,203 and FS90\,206 identify the
same object. 

\begin{table*}
\begin{minipage}{185mm}
\caption{FS90 galaxies in our MOSAIC field of the central Antlia region.} 
\setlength\tabcolsep{1.15mm}
\label{tabla}
\begin{tabular}{@{}ccccccccccr@{}c@{}lcr@{}c@{}lr@{}c@{}l@{}c@{}}
\hline
\multicolumn{1}{c}{FS90} &\multicolumn{1}{c}{NGC}  & \multicolumn{1}{c} {FS90 $\alpha$} & \multicolumn{1}{c} {FS90 $\delta$} & \multicolumn{1}{c}{FS90} & \multicolumn{1}{c}{FS90} &  \multicolumn{1}{c} {$E(B-V)$} &  \multicolumn{1}{c} {$T_1$} &  \multicolumn{1}{c}{$(C-T_1)$}  &  \multicolumn{1}{c} {$\mu_{_{T_1}}$}&  \multicolumn{3}{c} {$r_{_{T_1}}$}  &  \multicolumn{1}{c} {$<\mu_{\rm eff}>$} &  \multicolumn{3}{c} {$r_{\rm eff}$} & \multicolumn{3}{c}{v$_{\rm r}$} &  \multicolumn{1}{c}{Remarks}\\
\multicolumn{1}{c}{ID} &\multicolumn{1}{c}{}&\multicolumn{1}{c} {(2000)}&\multicolumn{1}{c} {(2000)} &\multicolumn{1}{c}{mor.} &\multicolumn{1}{c}{status} & \multicolumn{1}{c}{} & \multicolumn{1}{c}{\scriptsize mag}& \multicolumn{1}{c}{\scriptsize mag} & \multicolumn{1}{c}{\scriptsize mag arcsec$^{-2}$} & \multicolumn{3}{c}{\scriptsize arcsec} & \multicolumn{1}{c}{\scriptsize mag arcsec$^{-2}$} & \multicolumn{3}{c}{\scriptsize arcsec}  & \multicolumn{3}{c}{\scriptsize km s$^{-1}$} &\multicolumn{1}{c}{}\\
\hline
 68 &       & 10:28:03.1 & -35:26:31 & SBab        & 2 & 0.080 & 14.38 (0.01) & 1.60 (0.02) & 25.8 & 18&.&8 & 21.4 & 10&.&3 & 3189&$\pm$&80$^1$ &{\scriptsize  SE}\\
 69 &       & 10:28:05.0 & -35:28:55 & dE          & 2 & 0.081 & 18.88 (0.02) & 1.60 (0.04) & 25.7 &  7&.&5 & 23.3 &  3&.&1 &     &     &       &{\scriptsize  SE,DM}\\
 70 &       & 10:28:06.9 & -35:35:20 & dE          & 1 & 0.078 & 17.76 (0.02) & 1.64 (0.04) & 25.7 & 12&.&3 & 23.4 &  5&.&4 &     &     &       &{\scriptsize  SE}\\
 71 &       & 10:28:07.9 & -35:37:26 & Sd          & 1 & 0.076 & 16.31 (0.01) & 0.88 (0.02) & 25.8 & 13&.&6 & 22.3 &  6&.&2 &     &     &       &{\scriptsize  SE}\\
 72 &       & 10:28:07.9 & -35:38:20 & S0          & 1 & 0.075 & 14.39 (0.01) & 1.95 (0.02) & 25.7 & 16&.&9 & 20.4 &  6&.&4 & 3114&$\pm$&80$^1$ &{\scriptsize  SE}\\ 
    &       &            &           &             &   &       &        &      &       &   & &  &      &   & &  & 2986&$\pm$&38$^3$ &    \\ 
 73 &       & 10:28:09.8 & -35:43:04 & dE          & 1 & 0.073 & 16.97 (0.01) & 1.70 (0.02) & 25.7 & 10&.&5 & 22.2 &  4&.&4 &     &     &       &{\scriptsize  SE} \\
 75 &       & 10:28:12.0 & -35:32:20 & BCD?        & 3 & 0.081 & 17.65 (0.01) & 1.01 (0.02) & 25.7 &  4&.&0 & 20.7 &  1&.&6 &12450&$\pm$&95$^3$ &{\scriptsize  SE} \\
 76 &       & 10:28:12.9 & -35:35:38 & dE          & 2 & 0.079 & 18.04 (0.01) & 1.44 (0.02) & 25.8 &  5&.&4 & 22.2 &  2&.&7 & 25298&$\pm$&45$^3$ &{\scriptsize  SE}\\
 77 &       & 10:28:15.1 & -35:32:02 & dE,N pec or & 2 & 0.081 & 14.78 (0.01) &  1.80 (0.02) &  28.9 & 33&.&4 & 21.2 & 7&.&8 &  2382&$\pm$&49$^3$ &{\scriptsize  ELL,NS}\\
    &       &            &           & Amorphous?  &   &       &        &      &       &   & &  &      &   & &  &      &     &       &   \\
 78 &       & 10:28:15.8 & -35:46:26 & dE          & 1 & 0.076 & 19.36 (0.06) &  2.01 (0.06) & 27.4 &  10&.&8 & 24.5 &  4&.&3 &      &     &       &{\scriptsize  ELL}\\
 79 & 3258A & 10:28:19.2 & -35:27:21 & S0          & 1 & 0.082 & 13.20 (0.01) & 2.09 (0.02) & 27.9 & 50&.&7 & 19.6 &  7&.&5 &  2930&$\pm$&60$^1$ &{\scriptsize  ELL,NS}\\ 
    &       &            &           &             &   &       &        &      &       &   & &  &      &   & &  &  2734&$\pm$&36$^3$ &   \\ 
 80 &       & 10:28:18.9 & -35:45:28 & dS0         & 1 & 0.075 & 13.80 (0.01) & 2.08 (0.02) & 28.3 & 46&.&5 & 19.4 & 5&.&2 &  2544&$\pm$&80$^2$ &{\scriptsize  ELL,NS}  \\ 
 82 &       & 10:28:23.0 & -35:29:56 & S? or dS0? & 3 & 0.082 & 16.22 (0.02) &  2.34 (0.03) & 28.7 & 19&.&8 & 20.8 &  3&.&3 & 19577&$\pm$&41$^1$ &{\scriptsize  ELL,DM}\\ 
    &       &            &           &             &   &       &        &      &       &   & &  &      &   & &  & 19512&$\pm$&52$^3$ &   \\ 
 83 &       & 10:28:23.0 & -35:30:57 & S or Sm   & 3 & 0.082 & 16.31 (0.01) & 1.45 (0.02) & 25.7 & 11&.&9 & 21.8 &  5&.&0 &      &     &       &{\scriptsize  SE}  \\
 84 &       & 10:28:24.0 & -35:31:40 & E           & 2 & 0.082 & 13.70 (0.01) &  2.04 (0.02) & 28.3 & 42&.&3 & 19.4 & 5&.&4 &  2457&$\pm$&80$^2$ &{\scriptsize  ELL,NS}  \\
 85 &       & 10:28:24.0 & -35:34:22 & dE          & 1 & 0.081 & 18.12 (0.04) & 1.60 (0.07) & 25.6 & 14&.&6 & 23.9 &  5&.&6 &      &     &       &{\scriptsize  SE}  \\
 87 &       & 10:28:25.2 & -35:14:34 & dE,N          & 1 & 0.091 & 15.75 (0.01) & 1.87 (0.02) & 25.7 & 16&.&4 & 22.2 &  7&.&9 &      &     &       &{\scriptsize  SE,DM}\\
 88 &       & 10:28:28.0 & -35:31:04 & S or Sm  & 3 & 0.083 &  15.05 (0.01) &  1.79 (0.02) &  28.5 &  21&.&9 & 19.7 &   3&.&5 & 19659&$\pm$&80$^2$ &{\scriptsize  ELL}  \\
 93 &       & 10:28:31.9 & -35:40:40 & SmV         & 1 & 0.075 & 15.96 (0.01) & 0.91 (0.02) & 25.8 & 18&.&7 & 22.7 &  9&.&0 &      &     &       &{\scriptsize  SE}  \\
 94 &       & 10:28:31.9 & -35:42:21 & S0          & 1 & 0.074 & 13.22 (0.01) &  1.99 (0.02) & 28.6 & 44&.&0 & 17.9 &  3&.&5 &  2826&$\pm$&80$^2$ &{\scriptsize  ELL}  \\
 95 &       & 10:28:34.0 & -35:31:22 & dE          & 2 & 0.084 & 19.95 (0.04) & 1.57 (0.06) & 25.8 &  5&.&9 & 24.3 &  2&.&9 &      &     &       &{\scriptsize  SE}  \\
 98 &       & 10:28:35.0 & -35:27:39 & BCD         & 3 & 0.085 & 15.94 (0.01) & 1.25 (0.02) & 25.8 &  8&.&2 & 20.1 &  2&.&7 &  2890&$\pm$&94$^3$ &{\scriptsize  SE}  \\
103 &       & 10:28:45.1 & -35:34:40 & dE          & 3 & 0.084 & 19.95 (0.03) & 1.57 (0.04) & 25.8 &  4&.&9 & 23.8 &  2&.&3 &      &     &       &{\scriptsize  SE}  \\
105 & 3257  & 10:28:48.0 & -35:39:28 & SB01        & 1 & 0.077 & 12.77 (0.01) & 2.04 (0.02) & 28.2 &  69&.&4 & 18.6 &  5&.&8 &  3200&$\pm$&26$^1$ &{\scriptsize  ELL}  \\
106 &       & 10:28:51.3 & -35:09:39 & BCD?        & 3 & 0.098 & 17.10 (0.01) & 0.96 (0.02) & 25.8 &  8&.&7 & 21.6 &  3&.&1 &  2409&$\pm$&115$^3$&{\scriptsize  SE}  \\
108 &       & 10:28:53.2 & -35:19:12 & d:E,N       & 1 & 0.092 & 14.73 (0.02) &  2.03 (0.03) & 28.3 & 39&.&6 & 21.2 &  7&.&8 &  2611&$\pm$&39$^3$ &{\scriptsize  ELL,NS}  \\
109 &       & 10:28:53.0 & -35:32:52 & dE          & 2 & 0.087 & 18.99 (0.03) & 1.64 (0.05) & 28.0 &  7&.&6 & 23.1 &  2&.&6 &      &     &       &{\scriptsize  ELL,NS}  \\
110 &       & 10:28:53.0 & -35:35:34 & E(M32?)       & 3 & 0.085 & 15.49 (0.01) & 2.06 (0.02) & 27.5 & 14&.&0 & 18.4 &  1&.&5 &      &     &       &{\scriptsize  ELL}  \\
111 & 3258  & 10:28:54.0 & -35:36:21 & E           & 1 & 0.084 & 10.89 (0.01) &  2.18 (0.02) & 28.1 & 188&.&5 & 20.1 & 28&.&5 &  2792&$\pm$&28$^1$ &{\scriptsize  ELL}  \\ 
    &       &            &           &             &   &       &        &      &       &   & &  &      &   & &  &  2689&$\pm$&50$^3$ &   \\  
114 &       & 10:28:56.1 & -35:27:39 & dE          & 1 & 0.088 & 19.81 (0.06) & 1.45 (0.08) & 25.8 &  7&.&4 & 24.4 &  3&.&4 &      &     &       &{\scriptsize  SE}  \\
115 &       & 10:28:57.1 & -35:33:39 & dE          & 2 & 0.088 & 19.04 (0.04) & 1.64 (0.06) & 25.8 &  8&.&8 & 24.1 &  4&.&1 &      &     &       &{\scriptsize  SE}  \\
118 &       & 10:28:58.3 & -35:09:36 & dE          & 1 & 0.097 & 18.67 (0.06) & 1.71 (0.10) & 25.8 & 13&.&4 & 25.1 &  7&.&7 &      &     &       &{\scriptsize  SE}  \\
120 &       & 10:29:02.1 & -35:34:04 & ImV         & 1 & 0.088 & 16.38 (0.02) & 1.28 (0.02) & 25.8 & 17&.&4 & 22.8 &  7&.&6 &      &     &       &{\scriptsize  SE}  \\
121 &       & 10:29:02.1 & -35:35:34 & dE?         & 3 & 0.087 &   -    &  -   &  -    &   &-&  &   -  &   &-&  &      &  -  &       &{\scriptsize  DC}\\
123 &       & 10:29:03.1 & -35:40:30 & dE,N        & 2 & 0.080 & 16.33 (0.01) & 1.71 (0.02) & 25.8 & 12&.&5 & 21.9 &  5&.&1 &      &     &       &{\scriptsize  SE}  \\
125 & 3260  & 10:29:06.2 & -35:35:34 & S02         & 1 & 0.087 & 12.28 (0.01) &  2.12 (0.02) & 27.8 &  66&.&3 & 19.4 & 10&.&5 &  2416&$\pm$&32$^1$ &{\scriptsize  ELL,DM}\\ 
    &       &            &           &             &   &       &        &      &       &   & &  &      &   & &  &  2439&$\pm$&46$^3$ &   \\ 
131 &       & 10:29:11.0 & -35:41:24 & Sb(r)       & 3 & 0.081 & 14.26 (0.01) & 1.21 (0.02) & 25.8 & 16&.&4 & 20.5 &  7&.&1 &  2104&$\pm$&60$^3$ &{\scriptsize  SE}  \\
133 &       & 10:29:12.0 & -35:39:28 & d:E,N       & 1 & 0.083 & 14.63 (0.01) & 1.87 (0.02) & 25.8 & 16&.&5 & 20.6 &  6&.&4 &      &  -  &       &{\scriptsize  SE}  \\
134 &       & 10:29:13.2 & -35:29:24 & S0          & 2 & 0.089 & 14.23 (0.01) &  1.89 (0.02) & 27.2 & 40&.&0 & 19.7 &  4&.&9 &  1355&$\pm$&60$^3$ &{\scriptsize  ELL}  \\
136 &       & 10:29:15.3 & -35:25:58 & dE,N        & 1 & 0.090 & 16.12 (0.01) & 1.86 (0.02) & 25.8 & 14&.&5 & 21.9 &  5&.&9 &      &     &       &{\scriptsize  SE}  \\
137 &       & 10:29:15.1 & -35:41:34 & ImV         & 2 & 0.081 & 17.74 (0.02) & 0.81 (0.03) & 25.8 & 11&.&6 & 23.8 &  6&.&5 &      &     &       &{\scriptsize  SE}  \\
140 &       & 10:29:18.2 & -35:35:06 & dE,N        & 2 & 0.090 & 16.90 (0.02) & 1.88 (0.03) & 25.8 & 14&.&0 & 23.2 &  7&.&3 &      &     &       &{\scriptsize  SE}  \\
142 &       & 10:29:20.1 & -35:35:09 & dS0?        & 2 & 0.090 & 15.70 (0.01) & 1.85 (0.02) & 25.8 & 14&.&0 & 21.5 &  5&.&7 &      &     &       &{\scriptsize  SE}  \\
144 &       & 10:29:22.5 & -35:09:21 & dE          & 2 & 0.100 & 19.13 (0.04) & 1.56 (0.05) & 25.8 &  8&.&2 & 24.1 &  4&.&0 &      &     &       &{\scriptsize  SE}  \\
148 &       & 10:29:27.3 & -35:24:35 & dE/Im       & 3 & 0.093 &   -    &  -   &  -    &   &-&  &   -  &   &-&  &      &  -  &       &{\scriptsize  NS}\\
149 &       & 10:29:27.3 & -35:27:10 & S0 or dS0  & 3 & 0.091 & 16.95 (0.01) & 2.54 (0.02) & 25.8 &  5&.&7 & 20.5 &  2&.&1 &      &  -  &       &{\scriptsize  SE,DM}\\
153 &       & 10:29:31.4 & -35:15:39 & S0          & 1 & 0.101 & 13.24 (0.02) &  1.98 (0.03) &  28.4 &  50&.&4 & 18.7 & 4&.&9 &  1852&$\pm$&37$^1$ &{\scriptsize  ELL}  \\ 
    &       &            &           &             &   &       &        &      &       &   & &  &      &   & &  &  1733&$\pm$&39$^3$ &   \\ 
154 &       & 10:29:31.4 & -35:10:33 & dE/ImV      & 1 & 0.102 & 17.78 (0.10) & 1.92 (0.14) & 28.2 & 25&.&5 & 24.4 &  9&.&5 &      &     &       &{\scriptsize  ELL,NS}  \\
159 &       & 10:29:41.5 & -35:17:31 & d:E,N?      & 1 & 0.100 & 16.12 (0.01) & 1.80 (0.02) & 25.8 & 10&.&5 & 21.4 &  4&.&5 &      &     &       &{\scriptsize  SE}  \\
160 &       & 10:29:41.0 & -35:45:36 & dE          & 1 & 0.085 & 18.46 (0.05) & 1.73 (0.09) & 25.8 & 13&.&7 & 24.8 &  7&.&3 &      &     &       &{\scriptsize  SE}  \\
162 &       & 10:29:43.4 & -35:29:49 & dE,N        & 1 & 0.091 & 17.06 (0.02) & 1.75 (0.03) & 25.8 & 13&.&8 & 22.9 &  5&.&9 &      &     &       &{\scriptsize  SE}  \\
164 &       & 10:29:46.5 & -35:13:22 & ? or dE    & 3 & 0.104 & 18.37 (0.04) & 1.59 (0.06) & 25.8 & 12&.&3 & 24.2 &  5&.&8 &      &     &       &{\scriptsize  SE}  \\
165 &       & 10:29:46.0 & -35:42:25 & S0(M32?)      & 3 & 0.086 & 15.50 (0.01) & 2.01 (0.02) & 28.7 & 20&.&9 & 20.3 &  3&.&6 &  2605&$\pm$&80$^2$ &{\scriptsize  ELL}  \\
\hline
\end{tabular}
\end{minipage}
\end{table*}

\begin{table*}
\begin{minipage}{185mm}
\contcaption{}
\setlength\tabcolsep{1.15mm}
\begin{tabular}{@{}ccccccccccr@{}c@{}lcr@{}c@{}lr@{}c@{}l@{}c@{}}
\hline
\multicolumn{1}{c}{FS90} &\multicolumn{1}{c}{NGC}  & \multicolumn{1}{c} {FS90 $\alpha$} & \multicolumn{1}{c} {FS90 $\delta$} & \multicolumn{1}{c}{FS90} & \multicolumn{1}{c}{FS90} &  \multicolumn{1}{c} {$E(B-V)$} &  \multicolumn{1}{c} {$T_1$} &  \multicolumn{1}{c}{$(C-T_1)$}  &  \multicolumn{1}{c} {$\mu_{_{T_1}}$}  &  \multicolumn{3}{c} {$r_{_{T_1}}$} &  \multicolumn{1}{c} {$<\mu_{\rm eff}>$} &  \multicolumn{3}{c} {$r_{\rm eff}$} & \multicolumn{3}{c}{v$_{\rm r}$} &  \multicolumn{1}{c}{Remarks}\\
\multicolumn{1}{c}{ID} &\multicolumn{1}{c}{}&\multicolumn{1}{c} {(2000)}&\multicolumn{1}{c} {(2000)} &\multicolumn{1}{c}{mor.} &\multicolumn{1}{c}{status} &\multicolumn{1}{c}{} & \multicolumn{1}{c}{\scriptsize mag}& \multicolumn{1}{c}{\scriptsize mag} &  \multicolumn{1}{c}{\scriptsize mag arcsec$^{-2}$} & \multicolumn{3}{c}{\scriptsize arcsec} & \multicolumn{1}{c}{\scriptsize mag arcsec$^{-2}$} & \multicolumn{3}{c}{\scriptsize arcsec}  & \multicolumn{3}{c}{\scriptsize km s$^{-1}$} &\multicolumn{1}{c}{}\\
\hline
166 &       & 10:29:47.5 & -35:24:10 & E           & 2 & 0.097 & 15.28 (0.02)  & 2.35 (0.04) & 28.9 & 30&.&8 & 20.4   &  4&.&3  & 18658&$\pm$&99$^3$ & {\scriptsize ELL}  \\
168 & 3267  & 10:29:48.4 & -35:19:22 & SB01/2      & 1 & 0.100 & 12.37 (0.01)  & 2.16 (0.02) & 28.6 & 53&.&4 & 19.7   & 11&.&5   &  3709&$\pm$&33$^1$ & {\scriptsize ELL,DM}\\
    &       &            &           &             &   &       &         &      &       &   & &  &        &   & &    &  3773&$\pm$&65$^3$ &   \\  
169 &       & 10:29:48.4 & -35:25:12 & E           & 1 & 0.096 & -  & - & - & & -& &  -   &  & -&   &  3027&$\pm$&80$^2$ & \scriptsize GAP  \\ 
    &       &            &           &             &   &       &         &      &       &   & &  &        &   & &    &  2999&$\pm$&37$^3$ &   \\  
173 &       & 10:29:51.6 & -35:10:04 & d:E         & 1 & 0.105 & 13.93 (0.01)  & 2.04 (0.02) & 27.6 & 33&.&1 & 19.7   &  5&.&8   &  2650&$\pm$&80$^2$ & {\scriptsize ELL,NS}  \\ 
174 &       & 10:29:52.0 & -35:46:22 &  ? or dE,N  & 3 & 0.087 & 18.77 (0.02)  & 1.75 (0.03) & 25.8 &  6&.&2 & 23.1   &  2&.&9   &      &     &       & {\scriptsize SE} \\ 
175 &       & 10:29:53.5 & -35:22:37 & d:SB01      & 1 & 0.100 & 14.00 (0.01)  & 2.00 (0.02) & 28.5 & 45&.&0 & 20.5   &  8&.&1   &  1781&$\pm$&66$^1$ & {\scriptsize ELL}  \\
    &       &            &           &             &   &       &         &      &       &   & &  &        &   & &    &  1766&$\pm$&98$^3$ &   \\
176 &       & 10:29:54.4 & -35:17:16 & dE,N        & 1 & 0.102 & 17.24 (0.01)  & 1.71 (0.02) & 25.8 & 10&.&6 & 22.4   &  4&.&3   &      &     &       & {\scriptsize SE}  \\
177 &       & 10:29:54.4 & -35:19:19 & d:E,N       & 1 & 0.101 & 15.52 (0.01)  & 1.83 (0.02) & 25.8 & 12&.&1 & 21.1   &  5&.&2   &  3559&$\pm$&80$^2$ & {\scriptsize SE}  \\
    &       &            &           &             &   &       &         &      &       &   & &  &        &   & &    &  3505&$\pm$&45$^3$ &   \\ 
178 &       & 10:29:56.4 & -35:26:13 & dE          & 3 & 0.096 &   -     &  -   &-      &-  & &  &   -    &   &-&    &      &  -  &       & {\scriptsize DC}\\
179 &       & 10:29:56.4 & -35:31:40 & dE?         & 3 & 0.092 & 18.76 (0.01)  & 1.84 (0.02) & 25.8 &  4&.&4 & 22.6   &  2&.&4   & 56125&$\pm$&55$^3$ & {\scriptsize SE}  \\
182 &       & 10:29:57.1 & -35:42:46 & dE          & 3 & 0.088 & 18.77 (0.03)  & 1.65 (0.04) & 25.8 &  8&.&6 & 23.8   &  4&.&0   &      &     &       & {\scriptsize SE}  \\
184 & 3269  & 10:29:57.6 & -35:13:30 & S0/a        & 1 & 0.104 & 11.84 (0.01)  & 2.01 (0.02) & 28.8 & 84&.&5 & 19.4   & 13&.&0   &  3754&$\pm$&33$^1$ & {\scriptsize ELL,DM,NS}\\
    &       &            &           &             &   &       &         &      &       &   & &  &        &   & &    &  3753&$\pm$&99$^3$ &   \\ 
185 & 3268  & 10:29:58.5 & -35:19:30 & E           & 1 & 0.103 & 10.76 (0.01)  & 2.23 (0.02) & 27.1 & 192&.&3 & 20.7   & 39&.&5   &  2800&$\pm$&21$^1$ & {\scriptsize ELL}  \\
186 &       & 10:29:59.5 & -35:18:10 & dE          & 1 & 0.102 & 18.67 (0.02)  & 1.66 (0.03) & 25.8 &  7&.&4 & 23.6   &  3&.&9   &      &     &       & {\scriptsize SE}  \\
188 &       & 10:30:02.4 & -35:24:28 & dE          & 1 & 0.101 & 18.03 (0.02)  & 1.78 (0.03) & 25.8 &  8&.&4 & 23.2   &  4&.&4   &      &     &       & {\scriptsize SE,DM}\\
189 &       & 10:30:02.4 & -35:36:36 & dE/Im       & 2 & 0.094 & 18.31 (0.03)  & 1.86 (0.04) & 25.8 &  9&.&9 & 23.6   &  4&.&7   &      &     &       & {\scriptsize SE,DM}\\
192 &       & 10:30:04.5 & -35:20:31 & E(M32?)       & 3 & 0.104 & 16.72 (0.01)  & 2.11 (0.02) & 25.8 &  4&.&6 & 19.9   &  1&.&7   &      &     &       & {\scriptsize SE}  \\
193 &       & 10:30:04.3 & -35:32:52 & dE          & 2 & 0.093 & 19.33 (0.04)  & 1.37 (0.05) & 25.8 &  7&.&5 & 23.8   &  3&.&2   &      &     &       & {\scriptsize SE}  \\
195 &       & 10:30:06.4 & -35:18:25 & dE          & 1 & 0.104 & 19.25 (0.04)  & 1.62 (0.06) & 27.5 &  9&.&3 & 23.6   &  3&.&6   &      &     &       & {\scriptsize ELL,DM}\\
196 &       & 10:30:06.4 & -35:23:31 & dE          & 1 & 0.104 & 16.51 (0.01)  & 1.83 (0.02) & 25.6 & 12&.&7 & 22.4   &  5&.&9   &      &     &       & {\scriptsize ELL,DM}\\
201 &       & 10:30:13.6 & -35:15:54 & dE          & 1 & 0.103 & 18.68 (0.03)  & 1.74 (0.05) & 25.8 &  9&.&7 & 23.7   &  4&.&1   &      &     &       & {\scriptsize SE}   \\
202 &       & 10:30:15.3 & -35:27:32 & dE?         & 3 & 0.099 & 19.64 (0.03)  & 1.58 (0.04) & 25.8 &  5&.&5 & 23.7   &  2&.&6   &      &     &       & {\scriptsize SE}   \\
203 &       & 10:30:15.0 & -35:30:09 & d:E,N?      & 3 & 0.095 & 16.36 (0.01)  & 1.82 (0.02) & 25.8 & 10&.&8 & 21.1   &  3&.&6   &      &     &       & {\scriptsize SE,206}\\
205 &       & 10:30:18.4 & -35:24:43 & dE          & 2 & 0.105 & 17.71 (0.01)  & 2.02 (0.02) & 25.8 &  6&.&5 & 22.3   &  3&.&3   &      &     &       & {\scriptsize SE}  \\
207 &       & 10:30:18.4 & -35:31:26 & d:E,N       & 2 & 0.094 & 15.71 (0.01)  & 1.88 (0.02) & 25.8 & 10&.&6 & 20.8   &  4&.&1   &      &     &       & {\scriptsize SE}  \\
208 &       & 10:30:18.7 & -35:11:49 & S0(M32?)      & 1 & 0.103 & 14.85 (0.01)  & 1.95 (0.03) & 26.1 & 31&.&1 & 19.8   &  3&.&9   &  1774&$\pm$&100$^2$& {\scriptsize ELL}  \\
    &       &            &           &             &   &       &         &      &       &   & &  &        &   & &    &  1768&$\pm$&83$^3$ &   \\
209 &       & 10:30:19.4 & -35:34:48 & dE          & 2 & 0.096 & 18.09 (0.02)  & 1.72 (0.03) & 25.8 &  7&.&8 & 22.8   &  3&.&4   &      &     &       & {\scriptsize SE}  \\
212 &       & 10:30:21.3 & -35:35:31 & SmIII       & 1 & 0.096 & 15.58 (0.01)  & 1.31 (0.02) & 25.8 & 18&.&4 & 22.5   &  9&.&6   &      &     &       & {\scriptsize SE}  \\
213 &       & 10:30:21.6 & -35:12:14 & dE          & 2 & 0.102 & 19.16 (0.03)  & 1.66 (0.05) & 25.8 &  7&.&3 & 23.7   &  3&.&3   &      &     &       & {\scriptsize SE}  \\
214 &       & 10:30:22.5 & -35:30:32 & dE,N?       & 2 & 0.095 & 18.71 (0.03)  & 1.60 (0.04) & 25.8 &  8&.&2 & 23.5   &  3&.&7   &      &     &       & {\scriptsize SE,DM}\\
216 &       & 10:30:22.5 & -35:10:26 &  E     & 2 & 0.102 & 16.15 (0.01)  & 1.76 (0.02) & 25.8 &  7&.&8 & 20.3   &  2&.&7   &  2944&$\pm$&103$^3$& {\scriptsize SE}  \\
217 &       & 10:30:23.2 & -35:37:08 & dE?         & 3 & 0.097 & 19.98 (0.04)  & 0.95 (0.05) & 25.8 &  5&.&5 & 24.4   &  3&.&0   &      &     &       & {\scriptsize SE}  \\
220 &       & 10:30:24.7 & -35:15:18 & S0/a        & 1 & 0.102 & 14.16 (0.01)  & 1.85 (0.02) & 28.3 & 30&.&3 & 20.7   &  8&.& 1   &  1223&$\pm$&80$^2$ & {\scriptsize ELL,DM}\\
221 &       & 10:30:25.4 & -35:23:38 & dE          & 2 & 0.109 & 19.22 (0.04)  & 1.27 (0.05) & 25.8 &  8&.&2 & 24.1   &  3&.&8   &      &     &       & {\scriptsize SE}   \\
222 & 3258B & 10:30:25.4 & -35:33:43 & S0/a        & 2 & 0.095 & 14.15 (0.01)  & 1.88 (0.02) & 25.8 & 20&.&9 & 20.9   &  8&.&9   &  2140&$\pm$&80$^2$ & {\scriptsize SE}   \\
223 &       & 10:30:25.6 & -35:13:19 & dE,N          & 1 & 0.101 &   -     &  -   &     - &   &-&  &   -    &   &-&    &      &  -  &       & {\scriptsize BL}\\
224 & 3271  & 10:30:26.6 & -35:21:36 & SB02        & 1 & 0.108 & 11.13 (0.01)  & 2.25 (0.02) & 27.3 & 116&.&5 & 19.4   & 18&.&4   &  3737&$\pm$&27$^2$ & {\scriptsize ELL,DM}\\
    &       &            &           &             &   &       &         &      &       &   & &  &        &   & &    &  3803&$\pm$&31$^3$ &   \\
226 & 3273  & 10:30:29.2 & -35:36:36 & S0/a        & 1 & 0.097 & 11.84 (0.01)  & 2.23 (0.02) & 28.4 & 59&.&4 & 18.8   &  9&.&8   &  2419&$\pm$&52$^1$ & {\scriptsize ELL,DM}\\
227 &       & 10:30:31.4 & -35:23:06 & dE?         & 2 & 0.108 & 17.16 (0.01)  & 1.78 (0.02) & 25.8 &  8&.&6 & 22.0   &  3&.&8   &      &     &       & {\scriptsize SE}  \\
228 &       & 10:30:31.6 & -35:14:38 & dE,N        & 1 & 0.101 & 17.29 (0.02)  & 1.78 (0.03) & 25.8 & 12&.&6 & 22.8   &  5&.&1   &      &     &       & {\scriptsize SE}  \\
231 &       & 10:30:34.5 & -35:23:13 & d:E,N       & 1 & 0.108 & 14.98 (0.01)  & 2.06 (0.02) & 25.8 & 11&.&3 & 20.4   &  4&.&8   &  2931&$\pm$&80$^2$ & {\scriptsize SE}  \\
    &       &            &           &             &   &       &         &      &       &   & &  &        &   & &    &  2909&$\pm$&38$^3$ &   \\
235 &       & 10:30:39.6 & -35:31:44 & dE          & 3 & 0.097 &   -     &  -   &  -    &  -& &  &   -    &   &-&    &      &  -  &       & {\scriptsize DC}\\
237 &       & 10:30:44.6 & -35:19:12 & dE,N?       & 2 & 0.102 & 18.25 (0.02)  & 1.69 (0.03) & 25.8 &  8&.&3 & 23.1   &  3&.&6   &      &     &       & {\scriptsize SE,DM}\\
238 &       & 10:30:45.6 & -35:21:32 & Sm          & 1 & 0.104 & 15.45 (0.01)  & 1.47 (0.02) & 25.8 & 19&.&0 & 22.0   &  8&.&0   &      &     &       & {\scriptsize SE}  \\
239 &       & 10:30:47.5 & -35:28:55 & dE          & 2 & 0.101 & 18.99 (0.03)  & 1.52 (0.04) & 25.8 &  7&.&2 & 23.7   &  3&.&4   &      &     &       & {\scriptsize SE}  \\
241 &       & 10:30:48.4 & -35:32:20 & dE,N        & 1 & 0.098 & 16.77 (0.02)  & 1.84 (0.03) & 25.8 & 15&.&4 & 23.1   &  7&.&4   &      &     &       & {\scriptsize SE,DM}\\
\hline
\end{tabular}
\medskip

Notes.- Coordinates obtained through CDS, which are calculated from FS90. 
FS90 status 1 refers to definite members, status 2 to likely members, and 
status 3 to possible members. $\mu_{_{T_1}}$ corresponds to the threshold 
above which SExtractor detect and measures the object (MU\_THRESHOLD), or to 
the surface brightness of the outermost isophote for ELLIPSE. $r_{_{T_1}}$ is 
the radius that contains 90\% of the light for SExtractor. It is the 
equivalent radius ($r=\sqrt{a\cdot b}=a\cdot \sqrt{1-\epsilon}$) of the most 
external isophote for ELLIPSE. $\mu_{\rm eff}$ is obtained in both cases from 
$r_{\rm eff}$, the radius that contains one-half of the light. This radius is 
the output parameter HALF\_LIGHT\_RADIUS for SExtractor, and the 
equivalent effective radius for ELLIPSE. Radial velocities are from: 1= NED, 
2= 6dF, 3= our spectroscopic data. Remarks refers to: SE= Magnitudes and 
colours measured with SExtractor; ELL= Magnitudes and colours obtained from 
ELLIPSE; DM= Doubtful morphology, i.e. FS90 morphology does not match our 
images morphology; DC= Doubtful coordinates, i.e. FS90 coordinates do not 
clearly point to a galaxy; BL= bleeding; GAP= Within a gap of the image; 
NS= nearby bright star; 206 = also designated FS90 206.
\end{minipage}
\end{table*}

\section{Colours, magnitudes and surface brightnesses}
\label{photometry}

\subsection{The Colour-Magnitude Diagram}
\label{cm_diagram}

\begin{figure*}
\includegraphics[width=84mm]{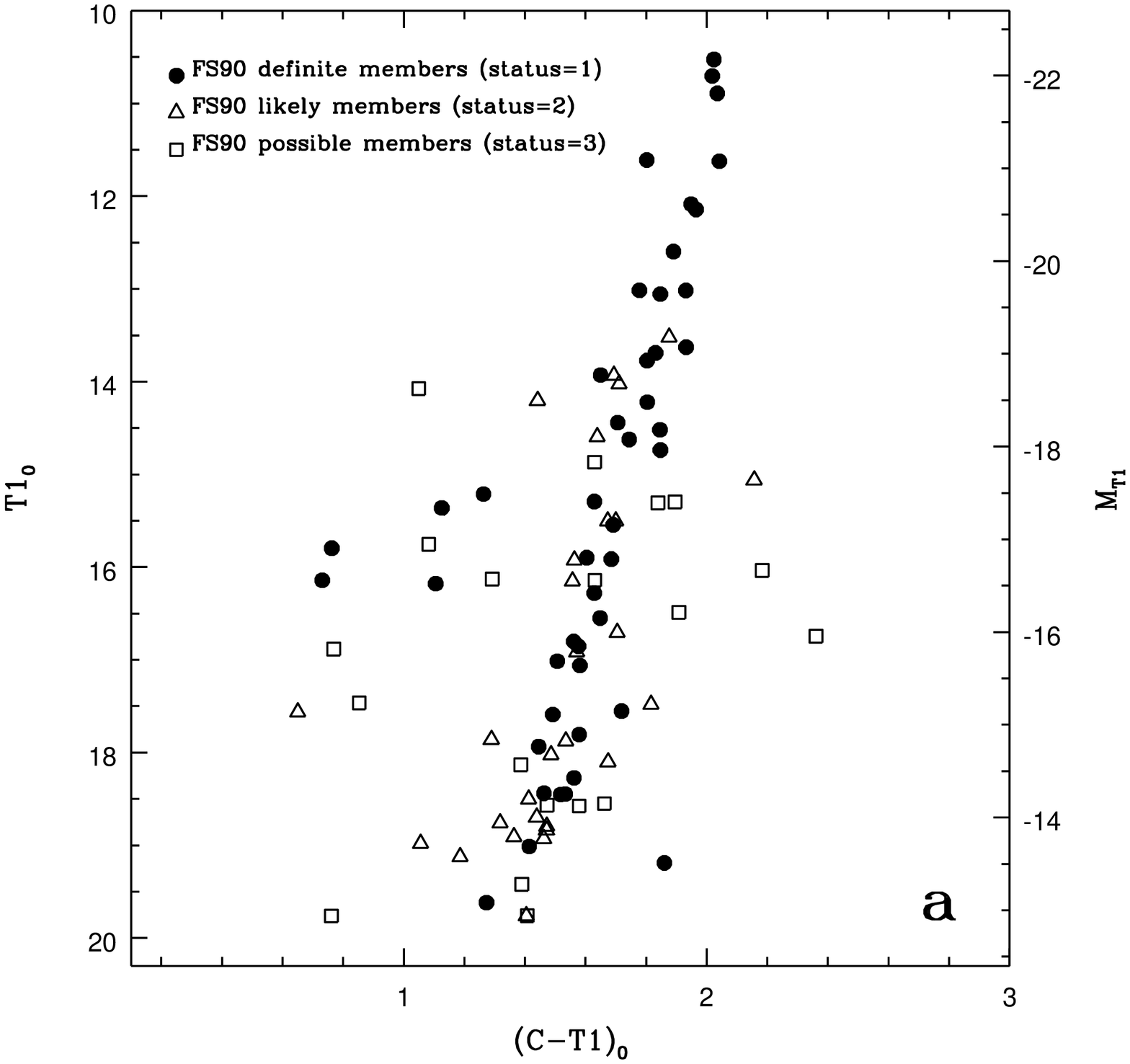}
\includegraphics[width=84mm]{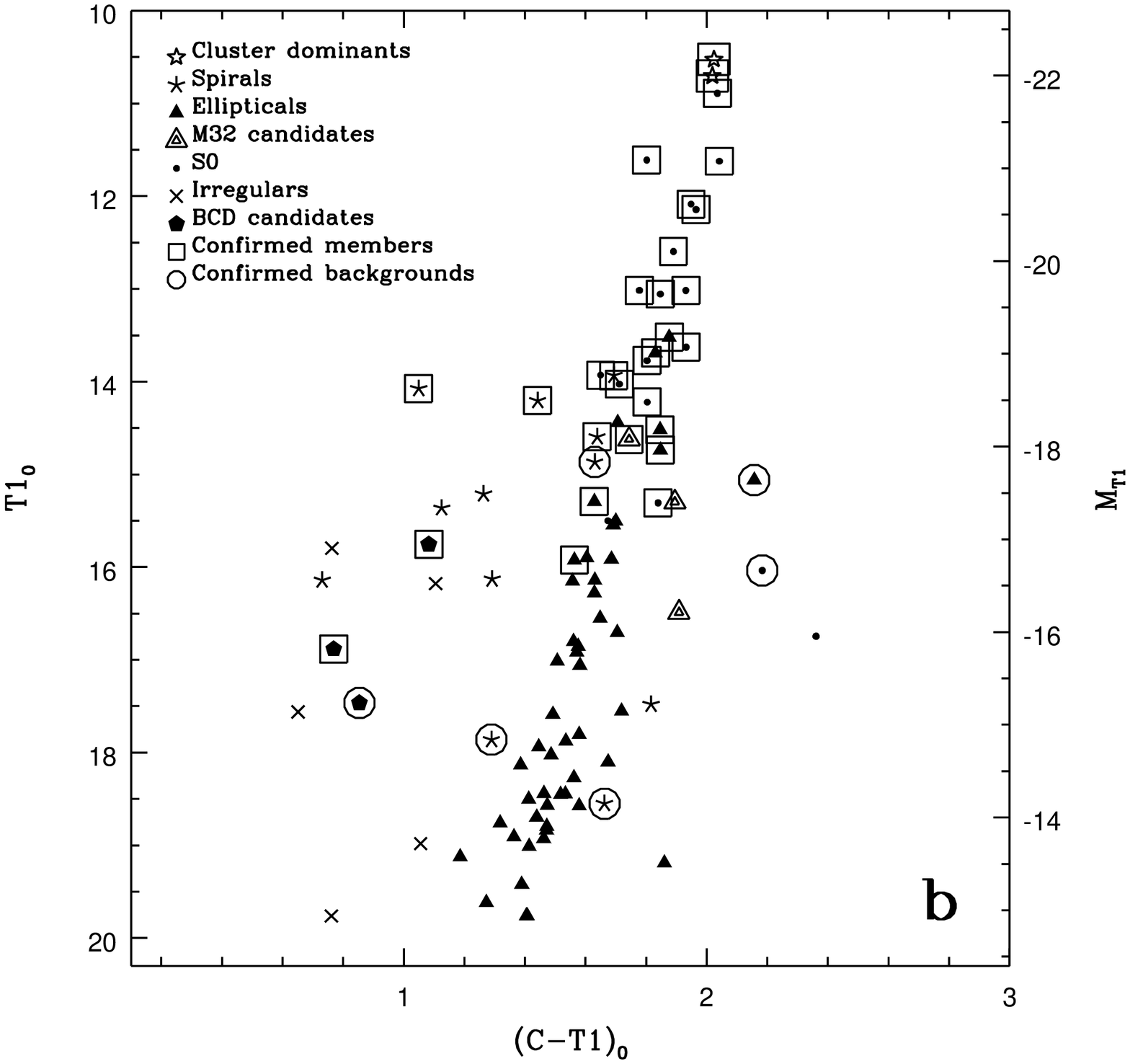}
\caption{CMD of 93 FS90 galaxies in our central field of Antlia. $T_1$ 
magnitudes and $(C-T_1)$ colours are absorption and reddening corrected. 
{\it Panel a:} the membership status according to FS90 is shown.
{\it Panel b:} small symbols indicate different morphologies, and large 
open symbols identify spectroscopically confirmed Antlia members and background 
objects.}
\label{CMD}
\end{figure*}

The left  panel of Fig.\,\ref{CMD} plots our galaxies from Table\,\ref{tabla} 
in the CMD according to their membership status given by FS90. Here the 
magnitudes and colours are reddening and absorption corrected
according to the relation $A_R/A_V=0.75$ \citep{RL85}. We got the $A_V$-values 
by looking up the individual reddening values for the galaxies \citep*{S98}  
and using the relation $A_V = 3~E(B-V)$. To transform $E(B-V)$ into $E(C-T1)$,
we applied $E(C-T1) = 1.97~E(B-V)$ \citep{HC77}. 

In the right panel of Fig.\,\ref{CMD} we show the same  galaxies, now 
indicating with different symbols their morphology. We distinguish 
spectroscopically confirmed Antlia members and background objects.
We visually inspected all our FS90 objects in order to see if their 
morphologies match with those given by FS90. Although there is a general 
agreement, in a handful of cases they display doubtful morphologies 
(see the remarks in Table\,\ref{tabla}). In these cases they are displayed in 
the plots with their FS90 morphological classification. 
In this first analysis we are classifying galaxies simply as spirals (S), Es, 
S0s, BCDs, and irregulars. A detailed morphological analysis will be given in 
a forthcoming paper.

In panel $a$ we can see that almost all FS90 definite members define
a quite narrow sequence. This sequence extends from the cluster dominant
galaxies to the dwarf regime, with no visible change of slope and no
increase in the scatter.  Five deviant objects lie towards bluer
colours. All are spirals or irregulars, supposedly star forming. Another
object at the faint end (FS90 78, $T_1\sim 19.2$, $(C-T_1)\sim 1.85$) 
shows a redder colour and is otherwise not
remarkable. It is a nucleated dwarf galaxy extremely faint on the
$C$-image and its colour might not be trustworthy.

From panel $b$, we see that practically all bright early-type galaxies
with membership status 1 are spectroscopically confirmed members. The only
elliptical which we spectroscopically confirmed as background galaxy, has membership 2 
in FS90 and clearly deviates from the mean relation.

Three galaxies classified as BCD are found at intermediate brightness in the
blue region. Two of them are spectroscopically confirmed Antlia members,
while the faintest one is a background object. All three were considered to
be status 3 members by FS90, given their morphologies. Our spectra,
dominated by strong emission lines, at least confirm that they are indeed
star forming.

In our field, FS90 also identified 4 objects as belonging to the very rare 
class of M32
type ellipticals \citep[e.g.][]{gra02,M05}. One of them, rather classified
as S0(M32?) (FS90\,165), seems to be an S0-galaxy and it is shown with this
morphology in Fig.\,\ref{CMD}. Another one (FS90\,208) is a confirmed
member, but does not deviate strikingly from the mean CM-relation. 
The other two are too red for their brightness, or too faint for their colour,
lying where background giant ellipticals should be found.
Although we have no radial velocities for them, their projected proximity to 
dominant Antlia galaxies, as well as their high surface brightness, lead us 
to consider them as M32-like candidates.

\begin{figure}
\includegraphics[width=84mm]{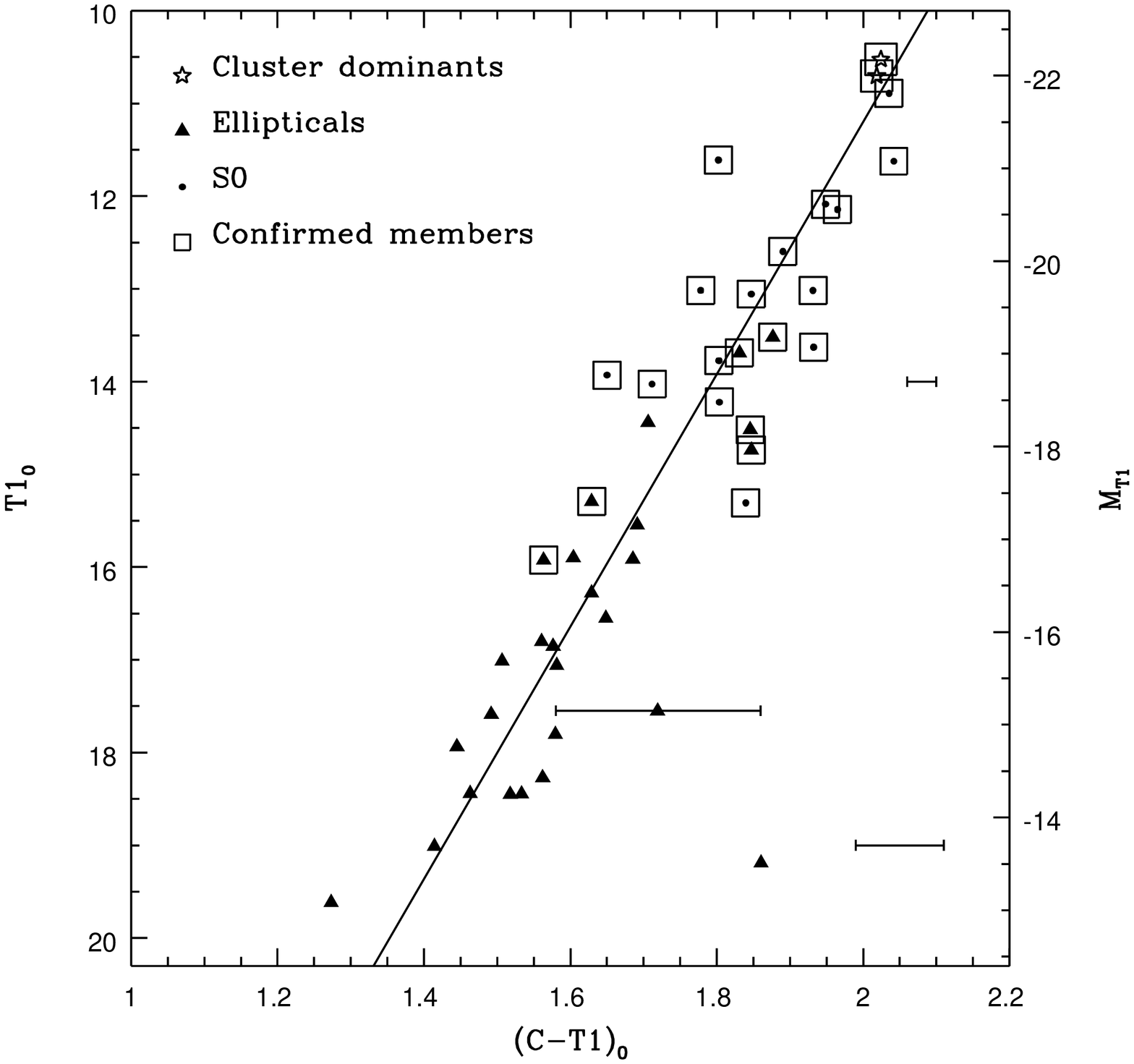}
\caption{This plot shows only confirmed members and dwarf galaxies with
a membership status of 1. The colour magnitude relation is tight for the 
fainter and broadens for brighter galaxies. S0s show bluer colours on the 
average. The isolated error bars show typical colour errors corresponding
to the range 10 $< T_1 <$ 18 (small) and 18 $< T_1 <$ 20 (large). We also 
display the error bar for the galaxy with the largest uncertainty.}
\label{status1}
\end{figure}

In Fig.\,\ref{status1} we show the positions of FS90 early-type galaxies (both 
E and S0) that are  considered definite members of Antlia (i.e. FS90 status 1 
early-type objects or galaxies that are spectroscopically confirmed members). 
Some galaxies are classified as SB0, and will be also included among the 
early-type galaxies. 

We see a  CMR that spans 9 magnitudes without a perceptible change of 
slope. The faintest galaxies show a tight CMR, while the individual deviations 
among the brightest galaxies can be substantial.
At the bright end, the S0-galaxies are slightly bluer on the average 
than the ellipticals ($\langle(C-T_1)_0\rangle_{S0}=1.91\pm0.09$,
$\langle(C-T_1)_0\rangle_{E}=1.94\pm0.10$) 
and display a larger dispersion about the mean relation.
Without spectroscopic information one cannot interpret this
difference in terms of age and/or metallicity (our medium-resolution spectra are
suitable only to obtain radial velocities). However, since S0s are
believed to be stripped spiral galaxies \citep[e.g.][]{D97}, an age 
difference is most likely. \citet{Ku00} has 
shown that, in the Fornax cluster, old stellar systems show tight scaling 
relations, and galaxies with young stellar populations tend to deviate from
these relations. In particular, he found that S0s display luminosity-weighted
ages less than those of Es, and show a considerable spread
about the scaling relations.

We performed several least-square fits to this relation, selecting the
subsamples listed in Table\,\ref{ajustes_CMR}. In all cases the fits were
performed considering the uncertainties in both coordinates \citep{NR}, and
rejecting the deviant faint object. From the fits we confirm that, within
the uncertainties, all brightest galaxies follow the same
relation as the rest of the early-type galaxies. 


\begin{table*}
\begin{minipage}{160mm}
\caption{Results of least-square fits $T_{1_0}=a+b\cdot(C-T_1)_0$ performed to 
the absorption and extinction corrected CMR of early-type definite members of 
Antlia (i.e. early-type status 1 objects and early-type galaxies 
with radial velocities). The first column indicates the different samples and 
the second column gives the number of data points. 
The limit magnitude to separate bright and dwarf galaxies ($T_1=14$ mag) 
corresponds to $M_V\sim -18$ mag \citep{G05}.}
\centering
\begin{tabular}{@{}lcr@{}c@{}lcc@{}}
\hline
\multicolumn{1}{c}{Sample} & Data & \multicolumn{3}{c}{$a$}  & $b$ & $\sigma_{(C-T_1)_0}$ \\
\hline
All definite members    & 43 & 38.4 & $\pm$ & 1.8   & $-13.6\pm1.0$ & 0.07 \\
Bright definite members & 15 & 44.6 & $\pm$ & 10.9  & $-16.8\pm5.0$ & 0.06 \\
Dwarf definite members  & 28 & 40.6 & $\pm$ & 4.7   & $-15.0\pm2.7$ & 0.08 \\
E definite members      & 28 & 37.8 & $\pm$ & 1.8   & $-13.3\pm1.0$ & 0.07 \\
S0 definite members     & 15 & 41.8 & $\pm$ & 10.7  & $-15.4\pm4.9$ & 0.09 \\
\hline
\end{tabular}
\label{ajustes_CMR}
\end{minipage}
\end{table*}

\subsection{The Surface Brightness-Luminosity Diagram}

Besides colours, relations between structural parameters of galaxies
can also tell us about their evolutive status, serving, at the same time,
to set membership criteria. For luminous ($M_B\lesssim -20$\,mag) E
galaxies, the effective radius $r_\mathrm{eff}$ tends to get larger while
the mean effective surface brightness $\langle \mu_\mathrm{eff}\rangle$
(i.e. the mean surface brightness within $r_\mathrm{eff}$) gets fainter
with increasing luminosity \citep[e.g.][]{K77}.

Early-type dwarfs, however, are known to display the opposite trends
\citep[e.g.][]{ferguson94}. This apparent dichotomy between low- and
high-luminosity ellipticals has been recently addressed by \citet{gra03},
who show that the general trend set by the fainter objects is broken by
the very brightest ellipticals, probably because of core formation.

\begin{figure*}
\includegraphics[width=84mm]{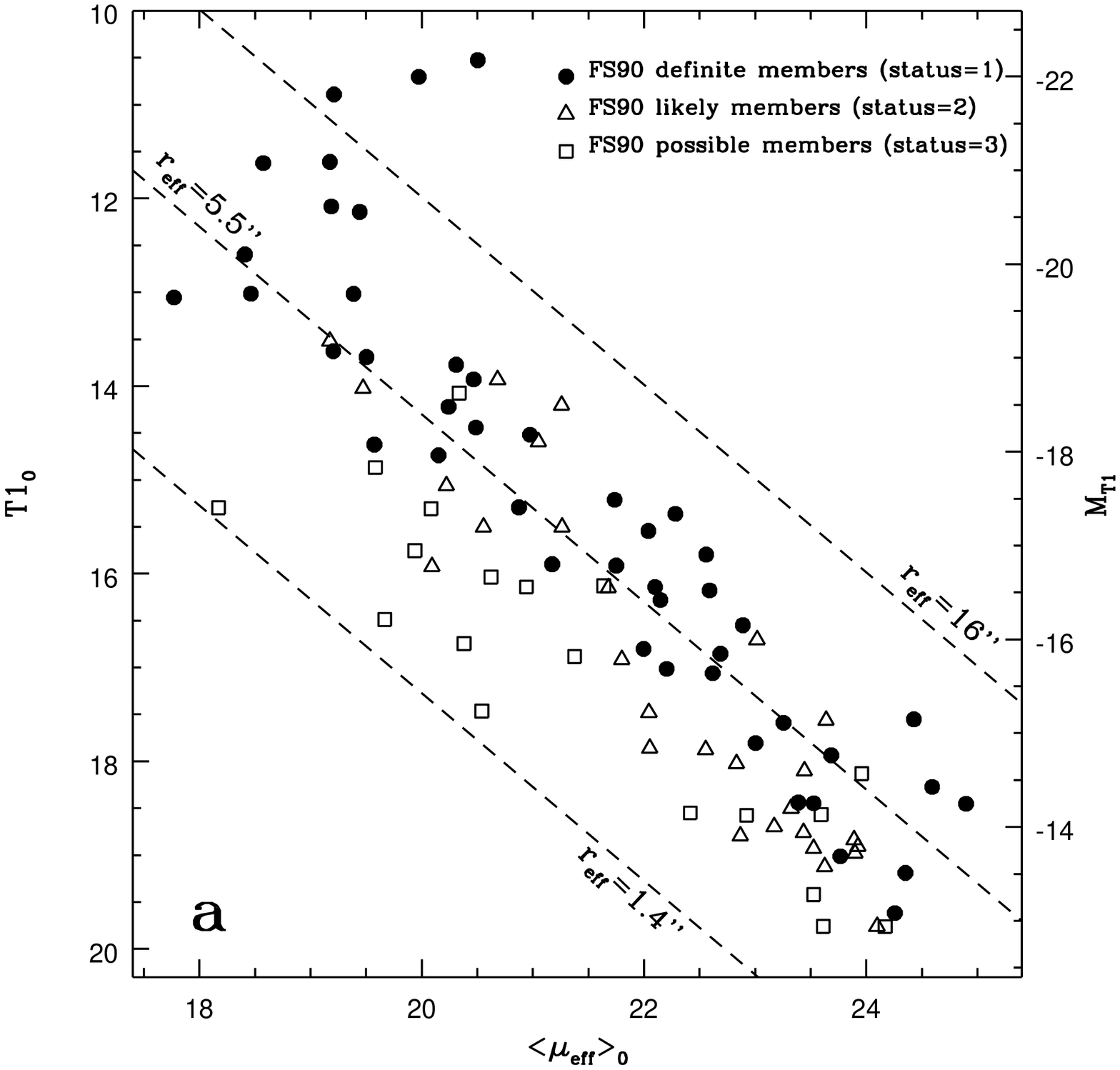}
\includegraphics[width=84mm]{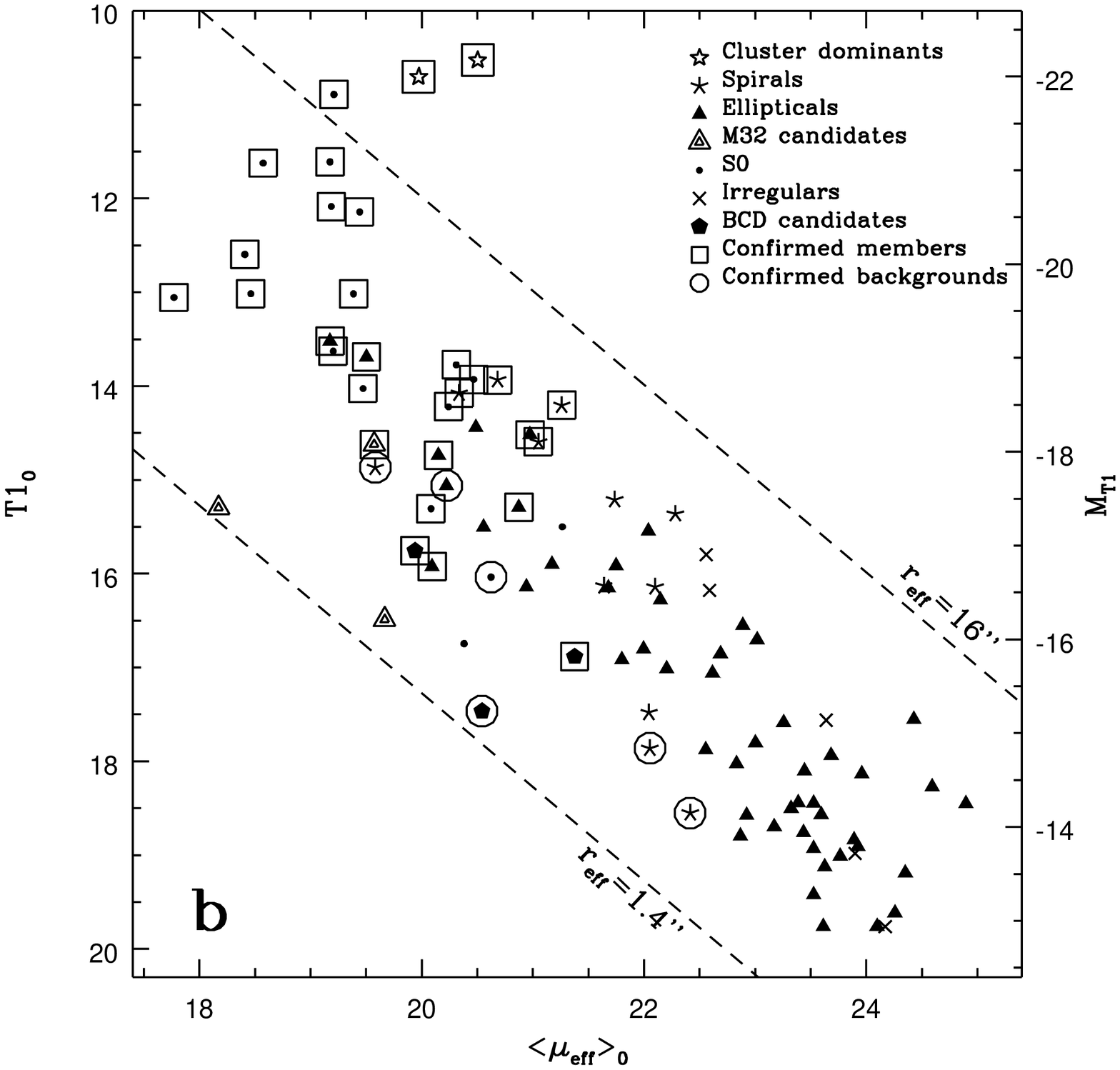}
\caption{Mean effective surface brightness vs. $T_1$ magnitude. {\it Panel a:}
  the galaxies according to their membership status. Shown are
  three lines of different effective radius, the middle line indicating the
  mean value for galaxies fainter than $T_1=13$\,mag. On the average, less
  likely members are displaced towards smaller effective radii. Status 2
  members scatter around a constant effective radius. {\it Panel b:}
  The different morphologies are indicated. Two
  M32-like candidates and one background Blue Compact Galaxy set the lowest 
  limit in effective radii of our sample.}
\label{Mueff}
\end{figure*}

Panel $a$ of Fig.\,\ref{Mueff} plots $\langle \mu_\mathrm{eff}\rangle$
versus $T_1$ for all Antlia FS90 galaxies with their membership status
indicated. As a reference we show two lines of constant effective radii,
corresponding to $1\farcs 4$ and $16''$, obtained from the definition of mean
effective surface brightness:

\begin{equation}
\langle \mu_\mathrm{eff}\rangle = T_1 + 2.5\,\log(2\,\pi\,r_\mathrm{eff}^2). 
\end{equation}

Status 1 galaxies in the range $13\lesssim T_1\lesssim 17.5$ mag 
($-19.7\lesssim M_{T_1} \lesssim -15$), nicely follow the relation for a 
constant effective radius of $5.5'' \pm 1.7''$.
At the assumed Antlia distance $1''$ subtends 170 pc \citep{dir03},
giving a mean effective radius of $0.94 \pm 0.3$ kpc.  
Although our sample is small in the
bright regime, there is a trend for the brightest galaxies to depart from
the general relation, consistently with \citet{gra03}.

Status 2 galaxies are found more or less in the same regime, but status 3
galaxies are offset to brighter $\mu_\mathrm{eff}$, i.e. to smaller
effective radii. This is consistent with the status assignment by
FS90, i.e. more compact objects are given a lower probability of being
members. Recently, \citet{CB05} have confirmed a large background contamination
among status 3 objects from FS90 in the NGC\,5044 Group, which is at a
similar distance as Antlia. A large fraction of background objects
should thus be expected among status 3 galaxies in Fig.\,\ref{Mueff}.

In panel $b$ of Fig.\,\ref{Mueff}, the morphological characteristics
are indicated. The confirmed background objects are consistently found at
lower $r_{\rm eff}$, as expected. Early-type and late-type galaxies
are not clearly separated, although the latter tend to show
more diffuse structures. Note also the location of the three galaxies 
classified by FS90 as M32-like dwarf ellipticals. While FS90\,208 follows the
same trend as the normal early-type galaxies, our M32-like candidates 
depart towards fainter magnitudes or higher surface brightnesses.

\section{Discussion}
\label{discussion}

\subsection{Surface brightness and effective radius}

Fig.\,\ref{Mueff} shows that the mean relation between $T_1$ and $\langle
\mu_\mathrm{eff}\rangle$ is to a good approximation the locus of a constant
mean $r_{\rm eff}$. In other words, the mean $r_\mathrm{eff}$ is largely
independent from luminosity. 
As a comparison, the Virgo dwarf galaxy sample by \citet{binggeli91}, for
example, exhibits the scaling law \citep[see also][]{ferguson94} $\langle
\mu_\mathrm{eff}\rangle = 0.75 \, M_B + 35.3$, meaning that $r_\mathrm{eff}$
grows with luminosity. To test if both relations are compatible, we 
applied a two-dimensional K-S test to the 
following datasets: the $\langle\mu_{\rm eff}\rangle$ of the Antlia dwarfs 
($T1>14$) transformed into the B band by means of the relations depicted 
in Sec 4.3, and the $\langle\mu_{\rm eff}\rangle$ obtained for the same 
galaxies but using the scaling law from Ferguson \& Binggeli (transforming 
the $M_{\rm T1}$ magnitudes into $M_B$). We 
calculated the statistic D \citep{NR} and got a probability 
$p=27$\%, which means that the hypothesis that these surface brightness vs. 
magnitude relations are different is not significant.  
Qualitatively similar trends can be seen in many other studies of
early-type dwarfs \citep[e.g.][]{CB87, VC94, C99, GTA04}. 

Our Antlia dEs, then, show a similar trend to the NGC\,5044 Group
sample of \citet{CB05}, where dwarf ellipticals with disk-like structure
tend to produce a slightly lower than unity slope. It will be interesting to
test whether examples of these candidates to harassed disk galaxies do exist
in the (presumably) dynamically younger Antlia cluster. A larger sample, 
as well as a careful evaluation of background contamination, incompleteness, 
and selection biases affecting the
$\langle \mu_\mathrm{eff}\rangle$ vs.\ luminosity relation will be necessary
to further study this issue.

\begin{figure}
\includegraphics[width=84mm]{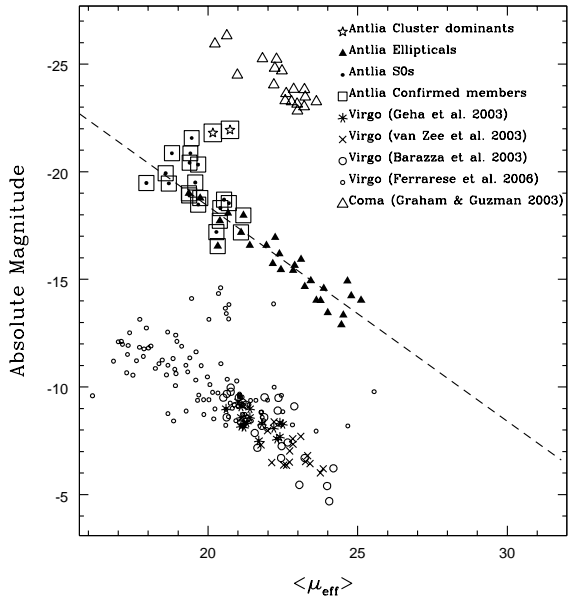}
\caption{Mean effective surface brightness vs. absolute magnitude for 
definite members of Antlia, and samples of Virgo and Coma galaxies. 
{\it The absolute mag scale is only valid for the Antlia galaxies}. The other 
samples are measured in other bands and most of them are shifted in mag to 
make the slopes better visible (the Virgo sample of \citet{Ge03} by adding 
8 mag, those of \citet{Bar03} and \citet{VZ04} by adding 9 mag, that from 
\citet{fer06} by adding 7 mag, and the Coma sample \citep{gra03} by subtracting
7 mag). The dashed line is the locus of a constant effective radius of the 
Antlia galaxies. The slope for Coma, and that of Virgo for galaxies 
fainter than $\sim - 13$ mag, are in agreement with a mean constant effective 
radius. The brightest galaxies from \citet{fer06} follow the perpendicular 
behaviour of the brightest Antlia members, typical of core galaxies 
\citep{gra03}.} 
\label{samples}
\end{figure}

In any case, our data are in principle consistent with a nearly constant
effective radius for dwarfs, and it is therefore interesting to compare our
mean $r_\mathrm{eff}$ value with CCD-photometry of recent samples in galaxy
clusters.
In Fig.\,\ref{samples} we plot data for dwarf galaxies in Virgo (\citealp*{Bar03,Ge03};
\citealp{VZ04}) and in Coma \citep{gra03}, where $\mu_{\rm eff}$ and $r_{\rm
eff}$ are obtained model-independently, and we also add the ACS Virgo sample
from \citet{fer06}. 
Since we are interested only in the slope of the relation between 
brightness and effective surface brightness,  
we just write {\it absolute magnitude} for the ordinate.
The straight dashed line corresponds to the mean effective radius of Antlia 
early-type definite members with $T_1 > 13$ mag. 
The individual samples use different bands and we did not try to homogenize 
them. 

Fig.\,\ref{samples} shows that all
samples are consistent with our slope, pointing to a constant mean 
$r_{\rm eff}$. 
The ACS sample also agrees if we consider galaxies
with absolute magnitudes fainter than $-13$ mag in our plot, except one 
deviant point 
at $-9.78$ in absolute magnitude. Note, however, that brighter
Virgo galaxies follow the perpendicular relation defined by the Antlia 
brightest members, already reported by \citet{gra03} for core galaxies. 
Considering distances of 17\,Mpc, 35.2\,Mpc, and 100\,Mpc for Virgo, Antlia, 
and Coma, respectively, we find the values listed in Table\,\ref{reff}. 
These values are in quite good agreement, although with 
substantial dispersions.

It is thus tempting to test the potential of a constant
$r_\mathrm{eff}$ as a distance indicator, in the light of previous efforts
in the same direction \citep[e.g.][and references therein]{BJ98, C99}.
However, other samples with a similar luminosity coverage are desired to arrive at
sensible conclusions.

\begin{table*}
\begin{minipage}{160mm}
\centering
\caption{Mean effective radius for 
the samples shown in Fig.\,\ref{samples}.
The fit to the ACS sample \citep{fer06} was performed
rejecting the deviant point at $-9.78$ in absolute magnitude.}
\begin{tabular}{@{}lccr@{}c@{}l@{}}
\hline
\multicolumn{1}{c}{Sample} & {Data} & {D} & \multicolumn{3}{c}{$\langle
  r_\mathrm{eff}\rangle$} \\
\multicolumn{1}{c}{} & & {Mpc} & \multicolumn{3}{c}{kpc} \\
\hline
Antlia definite members ($T_1>13$ mag)& 36 & 35.2 & 0.94&$\pm$&0.3\\
Coma \citep{gra03} & 18 & 100 & 0.97&$\pm$&0.3\\
Virgo \citep{Bar03}& 25 & ~~17 & 1.41&$\pm$&0.6\\
Virgo \citep{Ge03} & 17 & ~~17 & 0.96&$\pm$&0.2\\
Virgo \citep{VZ04} & 16 & ~~17 & 1.14&$\pm$&0.3\\
Virgo \citep[][$M_{\rm B}\gtrsim-20$]{fer06} & 87 & ~~17 & 1.26&$\pm$&0.8\\
\hline
\end{tabular}
\label{reff}
\end{minipage}
\end{table*}

\subsection{M32-like objects}

M32-like elliptical galaxies, which are distinguished by their low
luminosity, compactness, and high surface brightness, form a very rare
class. Although there are many candidates catalogued up to now (see for 
example, \citealp{B85}), beside M32 \citep[e.g.][]{gra02}, only five other 
objects are confirmed as 
such \citep[][and references therein]{Ch07}. FS90 classified thirteen 
galaxies as being M32-like in Antlia,
four of which are placed in our field.  As it was mentioned in
Sect.\,\ref{cm_diagram}, FS90\,165 seems to be an S0 galaxy. A radial
velocity is available only for the object FS90\,208, which according to its
surface brightness and effective radius seems to be a normal low luminosity
E galaxy. The other two are FS90\,110 and FS90\,192. 

It is noteworthy that, although their membership status has still to be 
settled, FS90\,110 and FS90\,192 lie close in projected distance to the two
dominant galaxies, and display high surface brightnesses. These facts are 
consistent with what is found for all confirmed M32-like elliptical galaxies, 
except the doubtful case of object 1 from \citet{M05}, which is 
about 5000 km s$^{-1}$ from the closest projected cluster giant 
elliptical. 
A detailed photometric analysis of Antlia M32 candidates will be given in a 
forthcoming paper.

\subsection{Colour-magnitude relation}

In order to compare our results with those reported in other papers, we
obtained transformation equations from Johnson - Cousins magnitudes and 
colours into the Washington photometric system. To do so, we used the results 
of the population synthesis models of \citet{B05}, who gives broad-band 
colours in several photometric systems for template galaxies of different ages 
spanning the whole Hubble sequence.  Linear relations with very low dispersions
were obtained for homologous colour indices [e.g. ($C-T_1$) vs. ($B-R$)] while
for colours probing different spectral regions [e.g. ($C-T_1$) vs. ($V-I_c$)]
linear fits had to be restricted to the corresponding sets of galaxy
types. These relations are:
\begin{equation}
V=T_1+(0.183\pm0.003)~(C-T_1)+(0.208\pm0.003),
\label{V_T1}
\end{equation}
\begin{equation}
(C-T_1)=(2.548\pm0.040)~(V-I_c)-(1.482\pm0.038),
\label{CT1_VI}
\end{equation}
\begin{equation}
(B-R)=(0.701\pm0.006)~(C-T_1)+(0.367\pm0.006).
\label{BR_CT1}
\end{equation}

By applying the SBF method and morphological 
classification to establish the membership status of Fornax dwarf galaxies, 
\citet{H03} and \citet{M07} have found a CMR for dEs 
with a scatter $\sigma_{(V-I)}=0.14$ ($\sigma_{(C-T1)}\sim 0.36$) 
down to $V=19$ mag ($T_1\sim18.6$ mag). In the Perseus cluster, 
early-type galaxies brighter than $M_B=-16$ mag ($T_1\approx$ 15.3 mag) 
display a tighter CMR with a scatter of $\sigma_{(B-R)}=0.07$ 
($\sigma_{(C-T_1)}\sim 0.1$), which is similar to our dispersion \citep{Con02}.
However, at $M_B=-13$ mag ($T_1\approx$ 18.5 mag), this scatter increases up 
to $\sigma_{(B-R)}=0.54$ ($\sigma_{(C-T_1)}\sim 0.77$), i.e., there is no 
longer a relation. 

\citet[hereafter LC04]{L04} have studied the CMR for early-type 
galaxies in 57 X-ray detected Abell clusters in the redshift range 
$0.02\le z \le 0.18$. They have found that the CMR is universal, with an 
average dispersion of $\sigma_{(B-R)}= 0.074 \pm 0.026 $
($\sigma_{(C-T_1)}\sim 0.1$). When these authors distinguish between low 
redshift and high redshift clusters, the mean dispersion turns to be 0.061 
for clusters with $z<0.04$, and 0.076 for the rest of the systems. 
Furthermore, LC04 noted that the cluster showing  the largest dispersion in 
its CMR (A2152, $\sigma_{(B-R)}=0.5$ at $R=18$ mag) belongs to the
set of systems that present background contamination from higher redshift
clusters (see fig.1 in their paper).

Regarding the slope of the relation in a $(C-T_1)$ vs. $T_1$ diagram, 
LC04 find a range between $-23.3$ and $-8.8$ with a trend of steepening at 
increasing redshifts. In particular, the Coma cluster CMR spans 8 mag in $R$, 
down to $R=21.2$, with a constant slope of $-15.2$. LC04 do not detect any 
significant change of slope within the CMR in any of their clusters. 

For Fornax, \citet{M07} derived a slope which is equivalent to -12.1 in our
CMD and, in the Perseus cluster, bright ellipticals follow a relation with a
slope of $-12.6$ \citep{Con02}. Furthermore, \citet{S97} find a CMR for the 
Coma cluster galaxies in the range $14 < R < 18.5$ mag, with a slope of 
-12.5 in a $T_1$ vs. $(C-T_1)$ diagram. Data from a previous Washington 
system study of Fornax dwarf elliptical galaxies by \citet*{CFG94}, 
although spanning a small magnitude range, are still consistent with our 
Antlia slope. 

\citet{LGB08} have recently obtained the CMR for dwarf early-type galaxies in 
the Virgo cluster, with SDSS data. In order to compare their slope with the 
one of our CMR, we transformed
our mean relation to the SDSS photometry system using equation\,(2) quoted 
above, and equations\,(4) and (7) from \citet*{J06}. We have obtained a 
value of $b=-0.022$ for a relation of the form $(g-r)=a+b\cdot r$. This is in 
good agreement with what is obtained by \citet{LGB08} for their full dE sample 
(excluding dEs with blue centres, dE(bc)). Our colour scatter transforms into 
$\sigma_{(g-r)}\sim 0.02$, which is smaller than those estimated by 
these authors.

From the previous analysis we can see that Antlia's CMR is one of the 
tightest and most extended relations reported up to now for nearby clusters, 
spanning a range of 9 magnitudes from cD to dwarf galaxies. 
Our $\sigma_{(C-T_1)}\sim 0.07$ is 
consistent with the scatter reported by LC04 for 57 X-ray detected clusters, 
and with that found for Perseus in its bright end. The large scatter reported
by \citet[][2003]{Con02} in the dwarf regime is probably due to background 
contamination as it was stated by LC04 \citep[see also][]{PC08}. Antlia's CMR  
displays no change of slope in agreement with other clusters, 
and its slope is consistent with those found in Fornax, Virgo, Perseus and 
Coma.

It is interesting to note that our CMR slope is also in agreement with 
that reported, in the Washington photometric system, for the metal-poor 
(blue) globular clusters ($M_{T1} > -10.4$, $M_V > -10.0$) associated to 
NGC\,4486 \citep*{FF07}. In this galaxy,
the mean colour of the globulars becomes redder with increasing luminosity, a 
behaviour that has been called `blue tilt' \citep[][and references therein]
{BS06}. Forte et al. have found a slope of $-16.67$. 

\subsection{Luminosity-metallicity relation?}

To transform colours to metallicity, we naively adopt the relation given 
by \citet{HH02} between $(C-T_1)$ and $\rm [Fe/H]$ derived for Galactic GCs. 
The transformation to our CMR then reads:
\begin{equation}
{\rm [Fe/H]} = -2.71 + \sqrt{-4.10 - M_{T1}/1.88}
\label{ec_Fe_T1}
\end{equation}
However, this relation describes old, single population objects.  A
significant fraction of dE galaxies is known to harbour young or
intermediate-age stellar populations \citep[e.g.,][]{CF96}, as well
as hidden discs, bars, spiral structure \citep[e.g.,][]{D01, BBJ02,
dRDZH03}, or even ongoing star formation at their centres
\citep[e.g.,][]{VSV84, CB01, LGWG06}. \citet{VZ04} found in their sample of
16 Virgo dwarfs that all galaxies are dominated by populations in the age
range 5-7 Gyrs. The mean age of 17 dwarfs was found by \citet{Ge03} to be 5
Gyrs. Therefore, the integrated $(C-T_1)$ colours of dwarf galaxies are
apparently determined by a mixture of age and metallicity, which we cannot
disentangle.

\begin{figure}
\includegraphics[width=84mm]{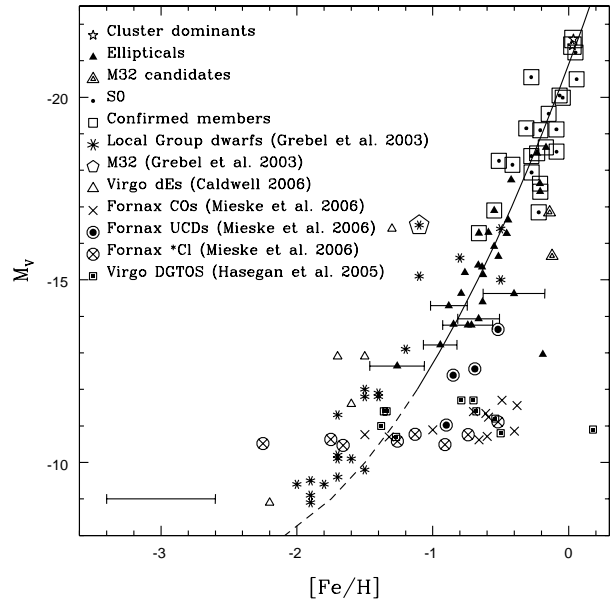}
\caption{$\rm [Fe/H]$ vs. $M_{\rm V}$ magnitude for different samples. 
As a reference, we show our mean $\rm [Fe/H]$ vs. $M_{\rm T1}$ relation as a 
solid line, as well as its extrapolation as a dashed line. Our typical error 
in $(C-T_1)$ (0.02 mag) translates into a $\rm [Fe/H]$ 
error of 0.03 dex by means of equation\,(\ref{ec_Fe_T1}). For clarity, we only
show error bars for Antlia galaxies that display uncertainties greater than
0.1 dex. The large isolated error bar corresponds to the mean metallicity error
in \citet{Gre03} sample (0.4 dex). } 
\label{Fe_Mv}
\end{figure}

Ignoring this complication, in Fig.\,\ref{Fe_Mv} we plot the luminosity -
metallicity relation of our Antlia galaxies.  In order to compare our
photometric metallicities with other samples, where mostly $M_V$ is given,
we transformed our $T_1$-magnitudes to $M_V$, using equation\,(\ref{V_T1}). 
Equation\,(\ref{ec_Fe_T1}) is shown as a 
reference with a solid line. We also plot Local Group dSphs from 
\citet*{Gre03}, Virgo dwarf-globular transition objects (DGTOs) from 
\citet{has05}, Virgo dEs from \citet{Cal06}, and the Fornax Compact Objects 
(FCO) from \citet{M06}.

The individual scatter of the published metallicities around the mean
relation is considerable and given the log-log character of this diagram,
more than a global statement of the kind that low-mass galaxies are
metal-poorer than high-mass galaxies is probably not permitted. It is more
interesting to note how tight the relation for Antlia galaxies is in all
its extension. 

Three of the five brightest galaxies of the Local Group sample  
(namely Sgr, NGC 185 and NGC 205) fall on the relation. 
The other two (M32 and NGC 147) depart towards higher luminosities
or lower metallicities, as fainter dSphs do. However, given the large 
metallicity errors of the Local Group dSphs
\citep[a typical value is 0.4 dex, see][]{Gre03}, we cannot rule out
that they follow the same trend as our mean luminosity-metallicity relation 
down to $M_V\sim-9$. 
Moreover, it
should be noticed that this mean relation is just an
extrapolation for magnitudes fainter than $M_V \sim -13$. In any case, 
irrespective of the curve defined by our mean relation in 
Fig.\,\ref{Fe_Mv}, dSphs seem to extend the luminosity- 
metallicity relation defined by the Antlia early-type galaxies, towards 
fainter magnitudes.

Virgo dEs seem to depart stronger from the Antlia mean relation than do the 
Local Group dSphs. Fornax and Virgo compact objects (COs and DGTOB), as well as 
Fornax bright globular clusters ($M_V<-10.4$, $\star$Cl) whose colour 
distribution is unimodal \citep[e.g.][]{O98}, do not obey any relation.  
Fornax UCDs seem to follow the galaxies' trend, although towards fainter 
magnitudes or higher metallicities. This might be 
pointing to the galaxy nature of UCDs.


Given all the cautious remarks on the limited applicability of our
photometric approach, Antlia galaxies define a tight 
luminosity-metallicity relation that extends over 9 magnitudes and 
might be followed by Local Group dSphs in its faint end. We note that 
these faint objects would not show this trend if we had 
used a linear relation between Washington colour and metallicity.

The astrophysical meaning of such a relation, which covers giant ellipticals, 
dwarfs, and likely dwarf spheroidals has been discussed extensively in the literature.
The formation history of these different types of galaxies is expected to
 be very different \citep[e.g.][]{derijke05}. Dwarf 
ellipticals probably were gas-rich late-type galaxies
which lost their gas by stripping or outflows related to
star formation activity \citep[e.g.][]{Ma05}. 
Dwarf spheroidals may also have a tidal origin.

Finally, we remark that the bright S0s outnumber the elliptical galaxies
roughly by a factor three. 
This is a difference to Fornax or Virgo, where most of the central galaxies are
ellipticals. If ellipticals form by the merging of disk galaxies while S0s
form by the gas removal of gas-rich disk galaxies, then merging was
apparently much less efficient in Antlia than close
encounters. Since S0s and ellipticals most certainly differ significantly in
their population composition, a common metallicity- luminosity relation is
not expected. S0s are thought to have a strong intermediate-age component,
which is in line with the separation between S0s and ellipticals in our
diagrams. We have no handle on the metallicity without spectroscopic
informations, but we may suspect that age or composite populations are
responsible for a large part of the scatter among the brightest galaxies.

\section{Conclusions}
\label{conclusions}

We conclude that spectroscopically confirmed or FS90 definite 
(i.e. status 1) early-type dwarf members of the
Antlia cluster define a very narrow sequence in the CMD. Luminous E and S0
galaxies follow the same CMR as their faint counterparts, with no
perceptible change in slope and with a slightly larger dispersion, which is due to
the separation of elliptical and S0 galaxies. 

This CMR spans 9 mag in brightness down to $T_1= 19.7$\,mag 
($M_B\approx -11.7$, $M_V\approx
-12.5$) with a small colour scatter of $\sigma_{(C-T_1)}\sim 0.07$\,mag.  
Our relation for only dwarf galaxies is tighter 
($\sigma_{(C-T_1)}\sim 0.08$\,mag) than that of most other samples found
in the literature. This may be due to 
our homogeneous data, obtained thanks to the good match of our MOSAIC field 
to the angular size of the central region of Antlia.  However, since
there is the possibility that colours of dwarf galaxies may be
influenced by younger stellar populations, one cannot be certain about
metallicities, and part of the scatter might still be due to different mean
ages.

The slope of the Antlia CMR is in agreement with those 
found in clusters like Fornax, Virgo, Perseus and Coma, despite of their 
different dynamical structure. This fact might indicate that the build up of 
the CMR is more related to internal galaxy processes than to the influence of 
the environment. Furthermore, the slope of this relation is also 
consistent with that displayed by blue globular clusters (`blue tilt') in 
NGC\,4486 \citep{FF07}. Previous comparisons between the `blue tilt' and the
positions of dE nuclei in the CMD, have already been performed by \citet{har06} and \citet{BS06}.
As this subject is out of the scope of the present paper, we plan to go on 
studying it in the near future.

We find a clear relation between luminosity and effective surface brightness
among the Antlia dwarf galaxies, scattering around a nearly constant
mean effective radius with a mild (if at all) dependence on
luminosity. A comparison with samples in the Virgo and Coma clusters
reveals consistency with a mean effective radius of about 1 kpc for some
samples. The dynamical meaning of this finding is unclear.

By applying the calibration between $(C-T_1)$ colours and $\rm [Fe/H]$ from 
\citet{HH02}, and interpreting the colour-magnitude relation as a 
metallicity-luminosity relation of old stellar systems, we find
that Antlia early-type galaxies seem to follow a tight luminosity-metallicity 
relation that extends from cD galaxies to the dwarf regime. Within  
metallicity uncertainties, Local Group dSphs might extend 
this non-linear relation covering a range of 13 mag. 
Fornax UCDs seem to define a luminosity-metallicity relation as well,
but towards fainter magnitudes or higher metallicities in comparison with
that displayed by Antlia early-type galaxies and Local Group dSphs.  This 
behaviour might point to a galaxy nature of UCDs.

The Antlia cluster provides a wealth of investigation possibilities which are 
still awaiting their exploitation. The present paper only refers to its central
region and to early-type galaxies. In forthcoming papers, we shall give  
brightness profiles, present new dwarf galaxies and perform spectroscopic 
studies of Antlia galaxies.
 We also plan to study the M32 candidates as well as the 
 UCD candidates that we might find in the Antlia fields.

\section*{Acknowledgements}

\textit{The measurement of new radial velocities has been done by Cristian
  Aruta.  We dedicate this paper to his memory.}

We thank the referee for a thorough reading of the manuscript and
for useful comments that helped to improve this paper.
We also thank Nicola Masetti for kindly providing observing time. This work was
funded with grants from Consejo Nacional de Investigaciones Cient\'{\i}ficas
y T\'ecnicas de la Rep\'ublica Argentina, Agencia Nacional de Promoci\'on
Cient\'{\i}fica Tecnol\'ogica and Universidad Nacional de La Plata
(Argentina). T.R. and L.I. are grateful for support from the Chilean Center for
Astrophysics, FONDAP No. 15010003. A.S.C. would like to thank Neil Nagar and
Universidad de Concepci\'on for their hospitality during her stay in Chile,
where part of this work was done.

\label{lastpage}
\end{document}